\documentclass[useAMS,usenatbib]{mnras}

\usepackage{times}
\usepackage{graphics,epsfig}
\usepackage{graphicx}
\usepackage{amsmath}
\usepackage{amssymb}
\usepackage{bm}
\usepackage{dcolumn}
\usepackage{epsfig}
\usepackage{subfig}
\usepackage{tabularx}
\usepackage[usenames,dvipsnames,svgnames,table]{xcolor}

\newcommand{\kms}{km~s$^{-1}$}

\newcommand{\ha}{\ensuremath{H\alpha~}}

\newcommand{\vn}{$\Delta v_{90}$}

\newcommand{\hi}{H{\sc i}\,\,}

\title[Mass and metallicity scaling relations]
{Mass and metallicity scaling relations of high redshift star-forming galaxies selected by GRBs
}
\author[Arabsalmani et al.]
{M. Arabsalmani$^{1,2,3,4}$\thanks{E-mail: maryam.arabsalmani@cea.fr}, P. M\o ller$^3$, D. A. Perley$^{4,5}$, W. Freudling$^3$, J. P. U. Fynbo$^4$, 
\newauthor
E. Le Floc'h$^1$, M. A. Zwaan$^3$, S. Schulze$^{6,7}$, N. R. Tanvir$^8$,   L. Christensen$^4$, 
\newauthor
A. J. Levan$^{9}$, P. Jakobsson$^{10}$,    D. Malesani$^4$, Z. Cano$^{11}$, S. Covino$^{12}$, V. D'Elia$^{13,14}$,     
\newauthor
P. Goldoni$^{15}$, A. Gomboc$^{16}$, K. E. Heintz$^4$, M. Sparre$^{4,17}$,  A. de Ugarte Postigo$^{4,10}$, 
\newauthor
S. D. Vergani$^{2}$ \\
        $^1$CEA/DSM/IRFU, CNRS, Universit\'e Paris-Diderot, 91190 Gif, France\\
        $^2$GEPI, Observatoire de Paris, PSL Research University, CNRS, Place Jules Janssen, 92190 Meudon, France\\
        $^3$European Southern Observatory, Karl-Schwarzschild-Strasse 2, 85748 Garching bei M\"{u}nchen, Germany\\
	$^4$Dark Cosmology Centre, Niels Bohr Institute, University of Copenhagen, Juliane Maries Vej 30, DK-2100 Copenhagen \O, Denmark\\
        $^5$Astrophysics Research Institute, Liverpool John Moores University, IC2, Liverpool Science Park, 146 Brownlow Hill, Liverpool L3 5RF, UK \\
        $^6$Instituto de Astrof\'i sica, Pontificia Universidad Cat\' olica de Chile, Vic\~una Mackenna 4860, 7820436 Macul, Santiago, Chile\\
        $^7$Millennium Institute of Astrophysics, Vic\~una Mackenna 4860, 7820436 Macul, Santiago, Chile\\
        $^8$Department of Physics and Astronomy, University of Leicester, University Road, Leicester, LE1 7RH, UK \\ 
        $^{9}$Department of Physics, University of Warwick, Coventry, CV4 7AL, UK\\
        $^{10}$Centre for Astrophysics and Cosmology, Science Institute, University of Iceland, Dunhagi 5, 107, Reykjavik, Iceland \\  
        $^{11}$Instituto de Astrof\'{\i}sica de Andaluc\'{\i}a (IAA-CSIC), Glorieta de la Astronom\'{\i}a s/n, E-18008, Granada, Spain\\
        $^{12}$INAF, Osservatorio Astronomico di Brera, Via E. Bianchi 46, I-23807 Merate (LC), Italy \\
        $^{13}${INAF-Osservatorio Astronomico di Roma, Via Frascati 33, I-00040 Monteporzio Catone, Italy}\\
        $^{14}${ASI-Science Data Centre, Via del Politecnico snc, I-00133 Rome, Italy}  \\
        $^{15}$APC, Astroparticule et Cosmologie, Universite Paris Diderot, CNRS/IN2P3, CEA/Irfu, Observatoire de Paris, \\ 
        Sorbonne Paris Cit\'E, 10, Rue Alice Domon et L\'Eonie Duquet, 75205, Paris Cedex 13, France\\
        $^{16}$School of Science, University of Nova Gorica, Vipavska 11c, 5270 Ajdov\v s\v cina, Slovenia\\   
        $^{17}$Heidelberger Institut f\"{u}r Theoretische Studien, Schloss-Wolfsbrunnenweg 35, 69118 Heidelberg, Germany\\      
}

\begin{document}
\date{}

\pagerange{\pageref{firstpage}--\pageref{lastpage}} \pubyear{}

\maketitle

\label{firstpage}

\begin{abstract}
{
We present a comprehensive study of the relations between   gas kinematics, metallicity, and stellar mass   in a   sample of 82  GRB-selected galaxies  using  absorption and emission methods. 
We find the velocity widths of both emission and absorption profiles to be a proxy of stellar mass. 
%Meanwhile, the velocity width of low-ion absorption lines appear to be several times larger than that of emission linesin GRB hosts. This could be due to the effect of interacting  systems or the significant contributions from galactic winds in broadening  the ISM absorption lines. If galactic winds dominate the velocity width of the ISM absorption lines, they appear to have much larger velocities in GRB host galaxies compared to the general star-forming galaxy population with similar stellar masses. This could be a result of the high star formation rate  densities in GRB hosts. 
We also investigate the velocity-metallicity correlation and its evolution with redshift  and find the correlation derived from emission lines to have a significantly smaller scatter compared to that found using absorption lines. Using 33 GRB hosts with  measured stellar mass and metallicitiy, we  study the mass-metallicity 
relation for GRB host galaxies {in a stellar mass range of $10^{8.2} M_{\odot}$ to $10^{11.1} M_{\odot}$ and  a redshift 
range of  $ z\sim 0.3-3.4$. 
The GRB-selected galaxies  appear to track the mass-metallicity relation of star forming galaxies but with an offset of 0.15   towards lower metallicities. 
This offset is comparable with  the average error-bar  on the metallicity measurements of the GRB sample and also 
the scatter on the MZ relation of the general population. It is hard to decide whether this relatively small offset is due to 
systematic effects  or the intrinsic nature of GRB hosts. }
%We show that GRB-selected galaxies  follow the mass-metallicity relation of the general star-forming galaxy population, in contradiction to several previous studies. In particular we show that   to within 3$\sigma$, GRB hosts  at most have metallicities 0.06 dex lower that that of emission selected  galaxies with similar stellar mass range.  
We also investigate the possibility of using absorption-line metallicity measurements of GRB hosts to study the mass-metallicity  relation at high redshifts. 
Our analysis shows  that the metallicity measurements from absorption methods can significantly differ from emission metallicities and assuming identical measurements from the two  methods may result in erroneous conclusions. }

\end{abstract}


%------------------------------------------------------------------------------------------------------------

\section{Introduction}
\label{sec:intro}

Long-duration Gamma Ray Bursts (GRBs) are beacons of star-forming galaxies \citep{Sokolov01-2001A&A...372..438S, Lefloch03-2003A&A...400..499L, Christensen04-2004A&A...425..913C, Fruchter06-2006Natur.441..463F} up 
to very high redshifts \citep[the highest confirmed spectroscopic redshift for a GRB is $ z=8.2$,][]{Tanvir09, Salvaterra09-2009Natur.461.1258S}. 
The detectability of these  extremely bright and dust-penetrating explosions is independent of the brightness and dust content of their host galaxies. 
Hence they provide a unique method for sampling star-forming galaxies  throughout the Universe without a luminosity bias, 
{something that significantly impacts even the deepest flux limited galaxy surveys.}  
%(as is the case for flux-limited galaxy surveys).  

The presence of GRB afterglows makes it possible to study their host galaxies  
through the absorption features that their interstellar media (ISM) imprint on the GRBs spectra \citep[see for e.g.][]{Castro03-2003ApJ...586..128C, Vreeswijk04-2004A&A...419..927V, Chen05-2005ApJ...634L..25C, Watson06-2006ApJ...652.1011W, Fynbo09-2009ApJS..185..526F, deUgartePostigo12-2012A&A...548A..11D} up to  the 
highest redshifts  \citep[][]{Sparre14-2014ApJ...785..150S, Hartoog15-2015A&A...580A.139H}.  
The fact that GRBs  fade away allow emission studies of their hosts  without interference of the bright GRBs 
\citetext{\citealp[for emission studied of GRB hosts see for e.g.][]{Savaglio09-2009ApJ...691..182S, Castroceron10-2010ApJ...721.1919C, Kruhler15-2015A&A...581A.125K, Perley16-2016ApJ...817....8P}}.
This is not the case for other absorbing systems such as those  in the sightlines of quasars where even at low redshifts detecting the 
galaxy counterparts have proven to be extremely  challenging due to  the presence of the bright background quasars \citep[][]{Warren01-2001MNRAS.326..759W, Christensen14-2014MNRAS.445..225C}. 
Independent measurements of galaxy properties (such as metallicity and gas kinematics)  using both absorption and emission methods 
and their connection with stellar mass 
can provide  insight into galaxy formation and evolution. GRBs provide an opportunity for performing such studies 
for a population of star-forming galaxies.

Absorption-line studies of GRB host galaxies   have led to accurate measurements  of abundances, metallicity,  dust, and kinematics  
up to redshifts $z\sim 6.0$ \citep[e.g.][]{Prochaska08-2008ApJ...672...59P, Fynbo09-2009ApJS..185..526F, Zafar11-2011A&A...532A.143Z, Thone13-2013MNRAS.428.3590T, 
Arabsalmani15-2015MNRAS.446..990A, Cucchiara15-2015ApJ...804...51C}. Emission-line studies have provided stellar masses, star formation rates, kinematics,  and emission-line metallicity 
measurements \citep[e.g.][]{Savaglio09-2009ApJ...691..182S, Castroceron10-2010ApJ...721.1919C, Kruhler15-2015A&A...581A.125K}, though limited to   redshifts   
$\lesssim 3.0$  due to the sensitivity limits of currently available telescopes {(also at $z \gtrsim 3$ the key diagnostic lines for emission-line metallicity measurements are redshifted out of the Near Infrared bands)}. 
However,  the connection between the information inferred  from the two methods  is yet to be studied. 
Metallicity measurements and kinematics are  two properties which are independently inferred from both 	 methods, using 
metal absorption profiles and bright nebular emission lines, respectively. 
These profiles   trace different regimes and  gas phases in galaxies. As a result, the two sets of line profiles typically 
have  different kinematic signatures \citep[e.g.][]{Castro-Tirado10-2010A&A...517A..61C}. Also, metallicity measurements from absorption 
and emission methods not only  trace the metal enrichment of gas in different regions of galaxies, but also are  based on totally different diagnostics 
\citep[see][]{Friis15-2015MNRAS.451..167F}. 

{
It is of much interest  to investigate whether GRB host galaxies  sample the  general star-forming galaxy population, or if 
they represent   a distinct galaxy population.  % This is crucial to before applying the findings from this selected population to the general population of galaxies.  
This question  has been the core of many studies in the research field of  GRB hosts \citep[see][as a few examples for addressing this question]{Fynbo08-2008ApJ...683..321F, Savaglio09-2009ApJ...691..182S, Arabsalmani15-2015MNRAS.446..990A, Greiner15-2015ApJ...809...76G, 
Schulze15-2015ApJ...808...73S, Perley16-2016ApJ...817....8P}. 
{Any systematic differences between GRB hosts and the
average field galaxy population provide important
clues as to the conditions required to produce GRBs,
and underpin attempts to use GRBs to probe galaxy evolution.
To date many works have indicated that GRB production
is disfavoured in high metallicity environments \citep[e.g,][]{Wolf07-2007MNRAS.375.1049W, Modjaz08-2008AJ....135.1136M, Savaglio09-2009ApJ...691..182S, Graham13-2013ApJ...774..119G, Vergani15-2015A&A...581A.102V, 
Perley16-2016ApJ...817....8P}, but it is less
clear whether other factors are also relevant 
as some GRBs have been associated with metal-rich galaxies \citep[e.g.,][]{Kruhler12-2012A&A...546A...8K, Savaglio12-2012MNRAS.420..627S}.  
A potentially
powerful diagnostic is the mass-metallicity relation, which
has frequently been discussed for GRB hosts \citep[][]{Savaglio06-2006AIPC..836..540S, Stanek06-2006AcA....56..333S, Kewley07-2007AJ....133..882K, 
Nuza07-2007MNRAS.375..665N, Levesque10-2010AJ....139..694L, Han10-2010A&A...514A..24H, Mannucci11-2011MNRAS.414.1263M, Graham13-2013ApJ...774..119G, Japelj16-2016A&A...590A.129J, Vergani17-2017A&A...599A.120V},
since it effectively allows us to investigate whether at
a given metallicity/redshift the hosts have typical masses.
%Studying the scaling relations   of   GRB host galaxies and comparing them to  those of  other galaxy populations would be the key to answer this question. The mass-metallicity relation has been frequently discussed for GRB host galaxies \citep[][]{Savaglio06-2006AIPC..836..540S, Stanek06-2006AcA....56..333S, Kewley07-2007AJ....133..882K, Nuza07-2007MNRAS.375..665N, Levesque10-2010AJ....139..694L, Mannucci11-2011MNRAS.414.1263M, Graham13-2013ApJ...774..119G, Japelj16-2016A&A...590A.129J}. 
However, the consistency of GRB host galaxies with the mass-metallicity  relation of the general star-forming galaxy
population has been  a subject under debate. }

{
In this paper we use a large sample of GRB host galaxies  with  measured properties from absorption and emission methods 
in order to combine 
%make a connection between 
our understanding of this galaxy population from the two 
methods. We study the scaling relations between gas kinematics, metallicity, and stellar mass and investigate 
their redshift evolution. }   
Our sample and the methods used to measure the galaxy properties are described in Section \ref{sec:sample}.   
In Section \ref{sec:VSM}  we compare the kinematic characteristics of gas in both emission and 
absorption, and investigate the connection between them  and the stellar mass. 
The relationships between gas kinematics and metallicity in both absorption and emission are discussed in Section \ref{sec:VZ}.   
Finally, we present the mass-metallicity relation 
for our large GRB host sample  in Section \ref{MZ} and compare it with that of the general population of star-forming galaxies. 
We discuss the possibility of using absorption metallicity measurements to study the mass-metallicity relation in Section \ref{ZX}. 
Our results are summarized in Section \ref{sec:sum}.

\begin{figure*}
\captionsetup[subfigure]{labelformat=empty}
\begin{center}$
\begin{array}{cc} \hskip -2mm
\subfloat[]
{\includegraphics[trim = 0mm 0mm 0mm 0mm, clip, width=0.50\textwidth, angle=0]{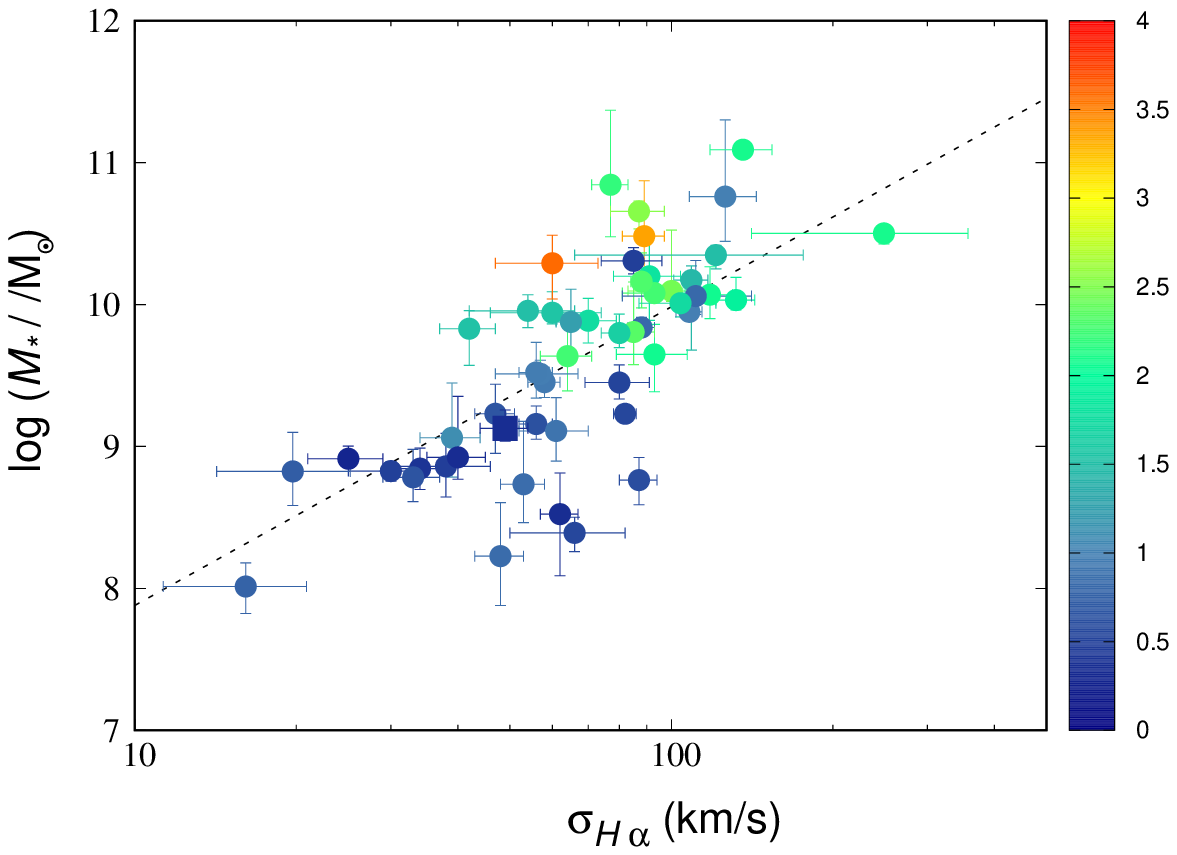}} 

& \hskip -5mm
\subfloat[]

{\includegraphics[trim = 0mm 0mm 0mm 0mm, clip, width=0.50\textwidth, angle=0]{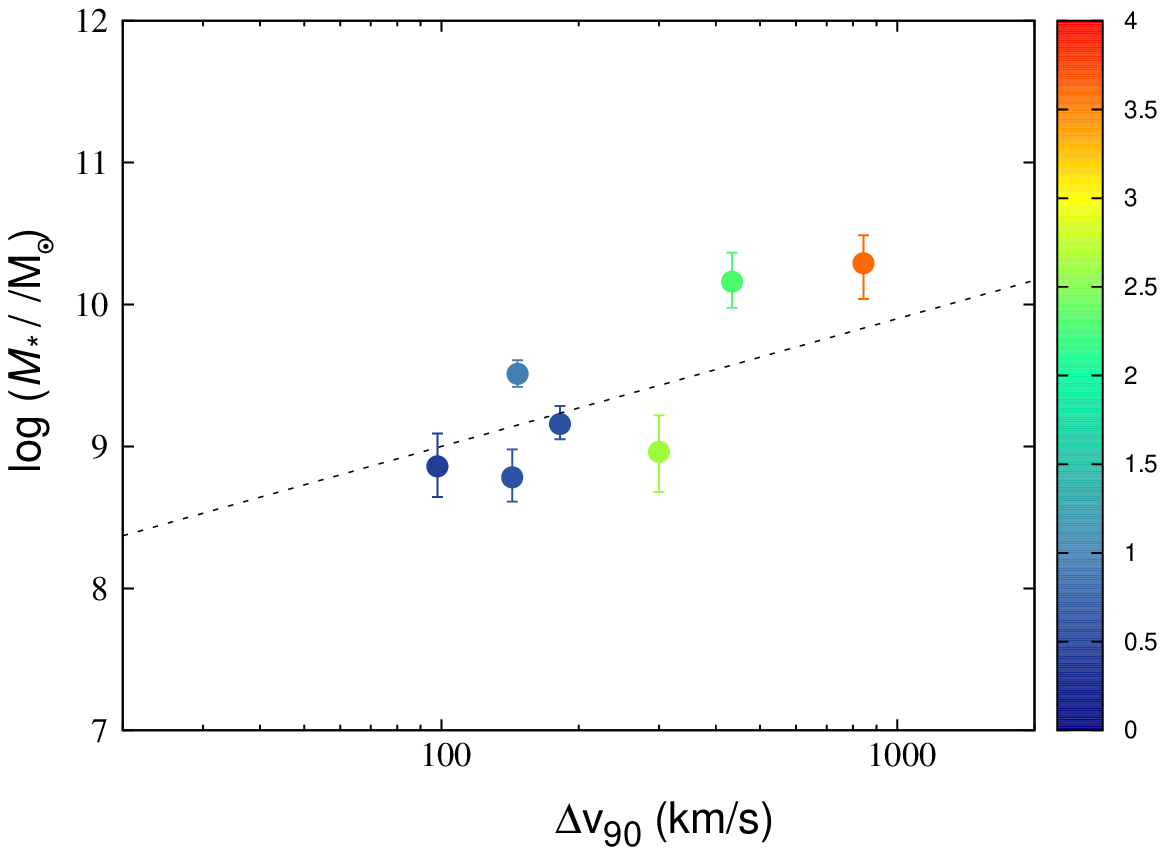}}
\end{array}$
\end{center}
\vskip -7.7 mm
\caption[Mass-velocity correlations]{\textit  {Left panel}: Stellar mass versus velocity width of bright emission lines, $\sigma_{H\alpha}$, for 52 GRB host galaxies that sample  a redshift range from $z=0.28$ to $z=3.58$. {The host galaxy of the ultra-long GRB 111209A is marked with a square. }
\textit  {Right panel}: Stellar mass versus velocity width of low-ion absorption lines, \vn, for seven  GRB host galaxies in our sample. The color-bars indicate the redshifts of the GRB hosts in both panels. The dotted lines  
show the best-fitting  lines obtained from the combination of data points 
in both plots and in Fig. \ref{fig:VS} (see Section \ref{sec:VSM} for details). 
}
\label{fig:MS}
\end{figure*}

\section[Sample and Measurements]{Sample and measurements}
\label{sec:sample}

\subsection{Sample}
{
Our main goal is to investigate the scaling relations  between the gas kinematics
and metallicity, inferred from both absorption and emission
methods, and  stellar mass for GRB host galaxies. In total we are then
considering five parameters describing five properties: absorption and
emission metallicities, absorption and emission velocity widths, and stellar
mass.  Currently all five parameters are known for only a single galaxy {(host of GRB 121024A)},
and we therefore compile (mainly from the literature)  a sample containing  GRB hosts for which at 
minimum two of the considered parameters are available. This allows us
to construct  sub-samples to study relations in any projection of the
5D parameter space. 
{In order to have a sample with consistently determined parameters we  use the 
sample of \citet[][]{Kruhler15-2015A&A...581A.125K} for  all the emission-line metallicity and  emission-line velocity width measurements. 
Also  all the stellar mass measurements are taken from  \citet[][]{Kruehler2017}.  
{
We take the absorption-line metallicities  from various  sources in the literature. 
Absorption-line velocity width measurements are either  presented in this work using the VLT/X-shooter GRB afterglow sample  (the  description 
of the data  is presented in Selsing et.al. in preparation) or 
taken from \citet[][]{Arabsalmani15-2015MNRAS.446..990A}.  
}

We do not  apply   any selection criteria based on the properties of GRBs themselves  
as such selection methods   do not necessarily imply well-defined selection criteria on properties of the hosts.
We therefore include the host galaxies of dust-obscured and/or dark GRBs \citep[GRBs with significant dust attenuation and/or $\beta_{OX}\gtrsim 0.5$; see][]{Jakobsson04-2004A&A...427..785J, Greiner11-2011AIPC.1358..121G, Perley13-2013ApJ...778..128P} in our sample.  
This %does not affect the scaling relations, but 
allows  us   to sample the largest possible range 
in the 5D parameter space which is critical in studying the scaling relations of GRB host population.} 
{ The two ultra-long GRBs, GRB 111209A  \citep[][]{Levan14-2014ApJ...781...13L} and GRB 130925A \citep[][]{Schady15-2015A&A...579A.126S},  are as well included in our sample. }

The full sample of 82 GRB host galaxies is listed in Table \ref{Tab:sample} {(see the on-line version of the paper for the complete table with all values 
listed)}. 
%where a reference is provided for each parameter which is known.
The relevant sub-samples count 
52 GRB host galaxies with stellar masses and emission-line velocity widths, 
43 with emission-line metallicities and emission-line  velocity widths, 
33 with stellar masses and emission-line metallicities, 
7 with stellar masses and absorption-line velocity widths, 
19 with absorption-line metallicities  and absorption-line velocity widths, 
3 with stellar masses and absorption-line metallicities, 
10 with emission-line and absorption-line velocity widths, 
and 1 GRB host  with emission-line and absorption-line metallicity measurements. 
%Note that we do not  imply  any selection criteria based on the properties of GRBs' prompt emission. This is justified for our study as we are interested in GRB hosts properties, and selection methods  based on GRBs' prompt emission properties do not necessarily imply well-defined selection criteria on properties of the hosts. This also allows having a large number of GRB hosts sampling large ranges of host properties which is critical in studying the scaling relations of GRB host population. 
}

\subsection{Measurements}  
\label{V}
{GRB host galaxies display very high column densities of neutral hydrogen, typically several times larger than 
the Damped Lyman-$\alpha$ (DLA) threshold \citep[see for e.g.,][]{Jakobsson06-2006A&A...460L..13J, Fynbo09-2009ApJS..185..526F}. In systems with such high \hi column 
densities the low-ion profiles trace the neutral hydrogen and hence the kinematic characteristics of these 
profile represent those of the neutral gas. The absorption profiles in GRBs spectra  usually show several components or clouds  tracing the velocity field in 
their host galaxies, similar to those of the  DLA systems  in the spectra of quasars. 
Each of these  clouds has a  broadening of a 
few \kms \citep[see e.g.,][]{Dessauges-Zavadsky06-2006A&A...445...93D}, but the total velocity width of the 
system is much larger, varying from a few tens of \kms \, to several hundreds of \kms. 
These absorption profiles trace only the gas in a narrow beam along 
the GRB sight-line and therefore the velocity width of these lines provide the  averaged velocity over the regions 
along the GRB sight-line only.}

To measure the velocity width of the neutral gas from low-ion absorption  profiles, we use  \vn\, as defined in \citet{Prochaska98-1998ApJ...507..113P}, 
which is the velocity interval that contains $90\%$ of the area under the apparent optical depth spectrum (see Fig. \ref{fig:v90} for an example). 
In order to measure \vn\, one needs to carefully choose the metal lines that are suitable for such measurement. Such a line should 
neither be weak nor saturated, as these would lead to under and over estimation of the velocity width, respectively \citep{Ledoux06-2006A&A...457...71L, Moller13-2013MNRAS.430.2680M}. 
Thus, we need to identify at least one low-ion metal profile in the GRB spectrum that is suitable for measuring the line-width. Identifying such a  
line for measuring the velocity width can be hard if the S/N of the spectrum is not high enough. 
In addition to this, we need to take care of the smearing effect caused by the resolution of the spectrograph.  
For this  we use the method discussed by \citet{Arabsalmani15-2015MNRAS.446..990A} 
and compute  the intrinsic velocity width  from 
\begin{eqnarray}
\Delta v_{90} = [\Delta v_{90,\rm meas}^2 - (1.4\times{\rm FWHM})^2]^{0.5}, 
\end{eqnarray}  
where $\Delta v_{90,\rm meas}$ is the measured value of the velocity width and FWHM is the corresponding Full-Width-at-Half-Maximum of the instrument resolution.  
{ The X-shooter spectrograph \citep[][]{Vernet11-2011A&A...536A.105V} consists of three separate spectrographs covering the spectral regions from the atmospheric cutoff to 550 nm (UVB), from 550 to 1000 nm (VIS) and from 1000 nm to the K-band (NIR). The spectral resolution is in the range 4000 - 17000 depending on the arm, the slit and/or the seeing during the observations. For the observations used in this study the FWHM of  spectroscopic resolution was typically in the range 30-60 \kms. }
%As discussed there, the correction can only be trusted if the resolution FWHM is not so large as to completely dominate over 
%the intrinsic width of the line. In case the intrinsic width is equal to the resolution then the measured width will be a factor
%of $\approx \sqrt{2}$ larger, which  is  fully  possible  to  correct  for.  
%We  therefore choose the same conservative approach 
%to only trust corrections if the measured width  is  less  than  $1.4$  times  the  width  after  correction.  In  otherwords, 
We use the $r$ parameter \citep[introduced in][]{Arabsalmani15-2015MNRAS.446..990A}, 
\begin{eqnarray}
\label{eq:r}
r := \frac{\Delta v_{90,\rm meas}-\Delta v_{90}}{\Delta v_{90}}
\end{eqnarray}
and choose a conservative approach  of only considering systems correctable if $r \leq 0.4$.  
% or equivalently  if $\Delta v_{90} \gtrsim {\rm FWHM}$.  \vn. 
We have X-shooter optical spectra  with sufficient S/N for  12 GRBs in order to measure  \vn\, (see Table   
\ref{Tab:v90}). The smearing effect of the instrument resolution does not allow  
\vn\,  measurements for two of them (see the values of parameter $r$ in  column 5 of Table  \ref{Tab:v90}). 

{ For gas seen in emission, we take  all the   \ha velocity dispersion  ($\sigma_{H\alpha}$) and the emission-line metallicity measurements  
from   \citet{Kruhler15-2015A&A...581A.125K} where they use VLT/X-Shooter observations of the host galaxies and base their  metallicity measurements  on 
calibrators from \citet[][]{Nagao06-2006A&A...459...85N} and  \citet[][]{Maiolino08-2008A&A...488..463M}. 
Stellar mass measurements are all taken from \citet[][]{Kruehler2017} where the measurements are based on  modeling  the Spectral Energy  Distributions  
(SEDs) of the hosts galaxies with LePhare \citep[][]{1999MNRAS.310..540A, 2006A&A...457..841I}, with galaxy templates from \citet{Bruzual03-2003MNRAS.344.1000B},  assuming exponentially declining star 
formation histories with the dust attenuation curve from \citet{Calzetti00-2000ApJ...533..682C}, and Chabrier IMF \citep[][]{2003PASP..115..763C}.  }

\section{Velocity width as a proxy of stellar mass}
\label{sec:VSM}

The relation between  gas kinematics and luminosity  was first introduced for nearby disk galaxies 
through the Tully-Fisher (TF) relation \citep[][using the inferred rotational velocity  from the \hi 21 cm emission line width]{Tully77-1977A&A....54..661T}. This was later extended to higher redshifts  
using optical lines, and to the relation between stellar mass and rotational velocity   
known as the stellar mass Tully-Fisher relation  \citep[sTF; see][for sTF at $0.1< z <1.2$]{Kassin07-2007ApJ...660L..35K}. Initial investigations of high redshift galaxies found no correlation \citep{Vogt96-1996ApJ...465L..15V, Simard98-1998ApJ...505...96S}, 
hinting  to anomalous kinematics of high redshift galaxies. This was confirmed by studies of Lyman Break galaxies at $z\sim3$ \citep{Pettini98-1998ApJ...508..539P, Pettini01-2001ApJ...554..981P} 
as well as UV-selected galaxies at $z\sim2$ \citep{Erb06-2006ApJ...644..813E}, from the integrated velocity width of nebular emission lines.  
However, recent studies (especially with the help of resolved 2D kinematics)  show that the sTF relation holds for high redshift galaxies, albeit with 
larger scatter compared to the local population \citep[][]{Puech08-2008A&A...484..173P, Puech10-2010A&A...510A..68P, Miller11-2011ApJ...741..115M, Glazebrook13-2013PASA...30...56G, Christensen17-2017MNRAS.470.2599C}.

We have stellar mass and $\sigma_{H\alpha}$ measurements    for 52 GRB hosts in our sample, covering 
 a redshift range from $z=0.28$ to $z=3.58$. The 52 hosts   are presented in the left panel of  Fig. \ref{fig:MS}. We clearly see a correlation 
between stellar mass and $\sigma_{H\alpha}$ \citep[see also][]{Christensen17-2017MNRAS.470.2599C}. 
The velocity width of the \ha emission line contains contributions from rotational velocity.  
But one should be careful not to erroneously interpret this width as an upper limit to the rotational velocity of the ionised gas 
in the galaxy. The full rotational velocity will only appear in the broadening of the \ha line if the observations are  
deep enough to pick up  faint emission from the full extent of the ionised gas in the star-forming disk. 
Therefore  we do not consider 
the  ${M_*}-\sigma_{H\alpha}$ relation, shown in Fig. \ref{fig:MS}, as a sTF relation. However,  the existence of such a correlation 
for  the GRB host sample with its large redshift range is interesting in   light of the sTF relation. 
{ We also emphasize that the velocity width of the \ha line should not be confused with the equivalent width of the \ha line. The latter 
measures the ratio of \ha flux, and hence Star Formation Rate  \citep[SFR,][]{Kennicutt12-2012ARA&A..50..531K},  to the stellar continuum. 
The \ha equivalent width thus  provides a proxy for  specific  SFR \citep[][]{Fumagalli12-2012ApJ...757L..22F}. Whereas the \ha velocity width   measures 
the velocity 
spread of the ionized gas. Therefore, the ${M_*}-\sigma_{H\alpha}$ relation  is not  directly representative   of the relation between 
stellar mass and SFR. }

We explore the existence  of a similar correlation between the stellar mass and \vn.  
We have stellar mass measurements for 7 GRB hosts with  \vn\, measurements.   
The right panel in Fig. \ref{fig:MS} shows these seven galaxies. Despite the small sample size, we can  
clearly see a trend of increasing stellar mass with increasing \vn.  
The two plots  in Fig. \ref{fig:MS} show that the velocity widths, measured from both absorption and emission methods, can be used as proxies for stellar 
mass. 
%The connection between the velocity widths and stellar mass  should not be surprising as all the components contributing to the velocity widths of both absorption and emission lines must be controlled by the gravitational potential in the galaxy.  

The relation between the velocity widths and stellar mass points to a correlation between the two velocity widths. We look for  the existence of
such a  correlation directly using the 10 host galaxies  for which we have   velocity width measurements in both emission and absorption ($\sigma_{H\alpha}$ and $\Delta v_{90}$, respectively). 
%At first instance, we clearly see that the values of $\sigma_{H\alpha}$ are several times smaller than $\Delta v_{90}$ values  for all 10 
Fig. \ref{fig:VS} shows the 10 GRB host galaxies in the $\Delta v_{90}-\sigma_{H\alpha}$ plane. 
As expected, we see a trend for an existing correlation between  the two velocity widths  (keeping in mind  the small sample size).

%But more interestingly, we see a trend for an existing correlation between  the two velocity widths  (keeping in mind  the small sample size). The relation between the two velocity widths points to the connection  between the velocity widths and mass. To investigate this further we explore the existence of such connections  directly using the stellar mass measurements of the GRB hosts in our sample. 

\begin{figure}
\begin{center} \hskip -5mm
\psfig{file=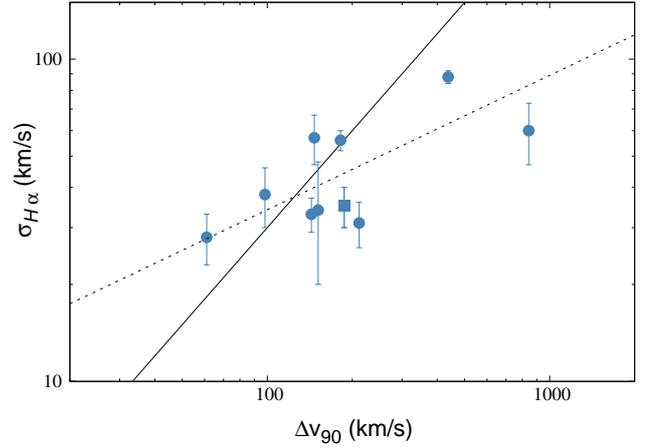,width=0.50\textwidth}
\end{center}
\caption[Emission vs. absorption velocity width]{The velocity width of low ion absorption lines, \vn,  versus 
the velocity width of  bright emission lines, $\sigma_{H\alpha}$, for 10 GRB host galaxies. The dotted line shows
the best-fit line when fitted together with the data presented in Fig. \ref{fig:MS} (see Section \ref{sec:VSM} for details). 
  The solid line ($\Delta v_{90} = 3.29\,\sigma_{H\alpha}$) shows the 
relation between the two velocities if the absorption and emission profiles were identical. {Note that the definition of the 
two velocity widths  are different from each other:  $\sigma_{H\alpha}$ is the 
standard deviation  of the fitted Gaussian function to the emission line while  \vn \, is  the velocity interval 
that contains $90\%$ of the area under the apparent optical depth. Hence, if the emission and absorption profiles were identical, 
the two  velocity widths would not be identical but   relate to each other as $\Delta v_{90} = 3.29\,\sigma_{H\alpha}$. 
{The host galaxy of the ultra-long GRB 111209A is marked with a square.}}   
}
\label{fig:VS}
\end{figure}

The three aforementioned relations, ${M_*}-\sigma_{H\alpha}$, ${ M_*}-\Delta v_{90}$, and 
$\sigma_{H\alpha}-\Delta v_{90}$, are not independent from each other. In order to quantitatively study 
these relations, we use the combination of all data points presented in  Figs. \ref{fig:MS} and  \ref{fig:VS}, and 
obtain the correlation parameters for the three relations simultaneously. 
This also allows us to have more reliable  results for ${M_*}-\Delta v_{90}$ and $\sigma_{H\alpha}-\Delta v_{90}$ 
relations where the sample sizes are small. We present our method  of finding the best fit correlation parameters 
using the combined data points in the Appendix. 
We find the two velocity widths (in \kms unit) to relate to stellar mass (is $M_{\odot}$ unit) as below:
\begin{eqnarray}
\label{eq:TF0}
&&{M_{*}} = 10^{5.8 \pm 0.4} \times \sigma_{H\alpha}^{2.1 \pm 0.2} ,
\end{eqnarray}

\begin{eqnarray}
\label{eq:TF}
&&{M_{*}} = 10^{7.2 \pm 0.7} \times \Delta v_{90} ^ {0.9 \pm 0.3} ,
\end{eqnarray}
with an intrinsic scatter of $0.4$ and $0.3$ dex in stellar mass for the two relations, respectively. 
And consistently, the two velocity widths follow the  relation:
\begin{eqnarray}
\label{eq:VS}
&&\sigma_{H\alpha}  \propto \Delta v_{90} ^{0.4 \pm 0.2}, 
\end{eqnarray}
with an intrinsic scatter of $0.1$ dex on $\rm\sigma_{H\alpha}$. %Excluding GRB 121219A reduces the scatter to 13.5 km/s.
The best fit correlations are shown as dotted lines in Figs. \ref{fig:MS} and \ref{fig:VS}. 
Note  that  the definition of the 
two velocity widths that we use are different from each other. The emission velocity width, $\sigma_{H\alpha}$, is the 
standard deviation  of the fitted Gaussian function to the emission line. But \vn \, is  the velocity interval 
that contains $90\%$ of the area under the apparent optical depth. 
For a Gaussian with a standard deviation of $\sigma$ the $90\%$ area is  between  
$-1.645\,\sigma$ and $+1.645\,\sigma$. Hence if the emission line profiles and the apparent optical depth of the absorption profiles  were identical,    
the two velocity widths should relate to each other simply as $\Delta v_{90} = 3.29\,\sigma_{H\alpha}$ (shown 
with a solid line in Fig. \ref{fig:VS}). This would only affect the intercept in the correlation shown 
in Fig. \ref{fig:VS} and would predict a slope of unity. Therefore the shallow 
slope of the correlation between the two velocity widths  (0.4 in  equation \ref{eq:VS}) is not an artefact of using 
differently defined velocity widths. 
{However, 
%all the points seem to be consistent with   the solid line but one, 
the GRB host with the largest \vn\, (host of GRB 090323A) in Fig. \ref{fig:VS} is the only point that is clearly inconsistent with the solid line.} To check the significance of the obtained correlation, we exclude this 
host from the sample and repeat the   fitting procedure. We do not find a significant change in 
the results, i.e. the slope of the correlation in Fig. \ref{fig:VS} remains well below one.

{
%Equation \ref{eq:TF0} is consistent with rotational motion being the dominating component in the broadening of the bright emission lines. \vn however appears to relate to stellar mass with a significantly larger power ($\Delta v_{90}\propto M_*^{1.1}$ compared to $\sigma_{H\alpha}\propto M_*^{0.48}$). This is translated to a shallow slope in Fig. \ref{fig:VS}. 
The  notably different powers in equations \ref{eq:TF0} and \ref{eq:TF}, or equivalently the shallow slope of the correlation shown in Fig. \ref{fig:VS}  
suggests significantly large \vn\, values especially for galaxies with large stellar masses.   \vn\, values larger than a few hundreds of \kms\, 
must have significant  contributions from components other than rotational motion. In the case of the host of GRB 090323A, the \vn\, of 843 \kms\, is significantly larger than the rotational velocity 
expected from its stellar mass of $M_* = 10^{10.3} \rm M_{\odot}$ {which is  213 \kms\, based on the  sTF relation presented in \citet[][]{Ubler17-2017arXiv170304321U}.} The situation is similar in the case of GRB 050820A host with a \vn\, 
of 300 \kms\, and an stellar mass of $M_* = 10^{8.96} \rm M_{\odot}$ (presented in the right panel of Fig. \ref{fig:MS}) 
{where the rotational velocity is expected to be  90 \kms \citep[based on the  sTF relation presented in][]{Ubler17-2017arXiv170304321U}.}

Large  contributions from galactic winds can  result in such large absorption widths.         
While rotation and random motions contribute to the broadening of both the emission and absorption  profiles, 
galactic winds  appear to primarily affect the width of the  absorption profiles.  
It is suggested that the \ha emission line is insensitive to a large fraction of the outflow mass, while the ISM absorption lines trace 
the global galactic winds \citep[e.g.][]{Wood15-2015MNRAS.452.2712W}. 
%The  significantly larger values of \vn\, compared to $\rm\sigma_{H\alpha}$ can be resulted from dominant contribution of outflows in absorption  line widths.  
However, the power-law index  of 0.9 in equation \ref{eq:TF} ($\Delta v_{90}\propto M_*^{1.1}$) %and comparing it with the findings for general star-forming galaxy population 
suggest larger galactic wind velocities in GRB host galaxies  compared to the general star-forming galaxy population \citep[see for e.g.][]{Arabsalmani17-2017arXiv170900424A}.
Several studies, based on both observations and simulation, have shown that in the general population 
of star-forming galaxies the outflow velocity relates to 
the stellar mass as $v_{\rm out}\propto M_{*}^{\sim 0.2}$ \citetext{\citealp[see e.g.][for observational study]{Bordoloi14-2014ApJ...794..130B, Karman14-2014A&A...565A...5K, Chisholm15-2015ApJ...811..149C}, \citealp[and][for studies based on simulation]{Barai15-2015MNRAS.447..266B}}.  
Also, the velocity of the infalling gas is expected to be smaller that the escape velocity and hence it should  
relate to stellar mass as $v_{\rm infall}\propto M_{*}^{\lesssim 0.3}$. 
Through simulations \citet{Lagos13-2013MNRAS.436.1787L}  show that  the outflow velocity increases  with the compactness of the star-forming region \citep[see also][]{Heckman15-2015ApJ...809..147H}. This should be  the case  in GRB host galaxies as they 	 have high SFR densities \citep{Kelly14-2014ApJ...789...23K} compared to the general galaxy population. %  with $v_{\rm out}\propto\rm M_{*}^{\sim0.2}$. 
This  is also supported by the presence  of  compact regions with recent star-forming activity   in GRB environments  seen in nearby GRB hosts 
\citetext{\citealp[see for example][for the host galaxy of GRB 980425]{Fynbo00}, \citealp[and][for the host of GRB 060505A]{Thone08-2008ApJ...676.1151T}},  
{ as well as GRBs  being coincident with the brightest regions in their host galaxies \citep[][]{Fruchter06-2006Natur.441..463F, Lyman17-2017MNRAS.467.1795L}}. 
%The larger power of $M_*$ in   equation \ref{eq:TF} (\vn $\propto M_*^{1.1}$)  %appears to be steeper a relation than  if \vn\, is a typical galactic wind velocity. could  therefore point to be expected for GRB hosts.  

 { 
Interacting systems and mergers, as with GRB 090323A
mentioned above,  could also result in large  absorption velocity widths. 
Indeed for the hosts of GRB 090323A and GRB 121024A, the two  host galaxies   
with the largest \vn,  %show  signature of interaction in their absorption profiles. In both systems, 
the absorption profiles contain two main components separated by a few hundreds \kms\, in velocity space, which could be due to two interacting  
galaxies \citetext{\citealp[for GRBs 090323 see][]{Savaglio12-2012MNRAS.420..627S}, \citealp[and for GRB 121024A see][]{Friis15-2015MNRAS.451..167F}}. 
In the case of GRB 050820A (mentioned above) \citet{Chen12-2012MNRAS.419.3039C}  proposed  the broad absorption signatures in the afterglow spectra 
to be due to the occurrence of the GRB in  a dwarf satellite  of an interacting system. 
Other evidence of interacting systems in GRB host galaxies have been  discussed by \citet{Chary02-2002ApJ...566..229C, Wainwright07-2007ApJ...657..367W, Chen12-2012MNRAS.419.3039C, Arabsalmani15-2015MNRAS.454L..51A}; and Roychowdhury et al. in preparation. %Our results confirm the possible connection between interacting systems and GRB events. }

}

\begin{figure}
\begin{center}
\psfig{file=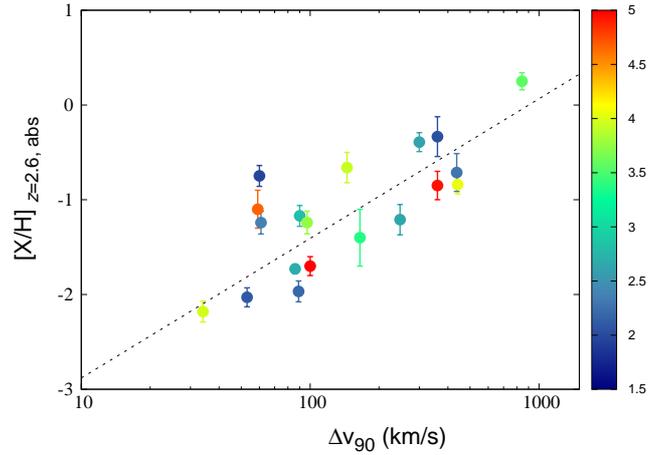,width=0.5\textwidth}
\end{center}
\caption[Velocity-metallicity correlation in absorption]{Correlation between absorption metallicity and velocity width of absorption lines, \vn,  
for 19 DLA systems  
intrinsic to GRB host galaxies. The metallicities are corrected for redshift and are 
set to the corresponding values at $z=2.6$ (see Section \ref{sec:VZ} for details). From this redshift onwards the correlation seems to 
remain unchanged. 
The color-bar indicates the redshifts of the GRB hosts.  
}
\label{fig:v-z}
\end{figure}

\section{Velocity-metallicity correlation in both absorption and emission}
\label{sec:VZ}

Previous studies have shown that the velocity width of low ion absorption lines correlates linearly with the metallicity (inferred 
from absorption lines) for Damped Lyman-$\alpha$ galaxies   in the sightlines of quasars \citep{Ledoux06-2006A&A...457...71L, Prochaska08-2008ApJ...672...59P, Moller13-2013MNRAS.430.2680M, Neeleman13-2013ApJ...769...54N}. 
Moreover, \citet{Moller13-2013MNRAS.430.2680M} found that the velocity-metallicity (VZ) correlation  evolves linearly with redshift up to $z=2.6$ and 
then remains unchanged  for $z>2.6$. This correlation is proposed to  be representative of a mass-metallicity (MZ) relation for this population of high redshift 
galaxies. \citet{Christensen14-2014MNRAS.445..225C} confirmed the consistency of the VZ correlation with the MZ relation for a sample 
of 12 DLA galaxies with measured stellar masses. 
Stellar mass measurements for DLA systems in sightlines of quasars  have proven to be extremely 
challenging (especially due to the presence of the bright background quasar).

\citet{Arabsalmani15-2015MNRAS.446..990A}  performed the same study  
for the DLA systems that are intrinsic to GRB host galaxies  and concluded that GRB-DLAs not only follow a VZ correlation, but they are also 
consistent with  that of QSO-DLAs \citep[see also][]{Prochaska08-2008ApJ...672...59P}. They also found 
the VZ correlation of GRB-DLAs to obey the same redshift evolution as  QSO-DLAs. % {\citep[see][for a detailed discussion on the redshift evolution of the VZ correlation]{Moller13-2013MNRAS.430.2680M}}.  
Fig. \ref{fig:v-z} shows the VZ correlation for  19 GRB hosts  \citep[16 of them are  presented in][]{Arabsalmani15-2015MNRAS.446..990A} with 
the host metallicities  shifted to the corresponding metallicities at $z=2.6$  using the evolution of the VZ correlation derived by \citet{Moller13-2013MNRAS.430.2680M}. {In order to shift the metallicity of each host to a reference redshift (here $z=2.6$) we calculate the offset between the 
measured  metallicity  and the VZ correlation at the redshift 
of the host. We  then place the host at  the same  offset from the VZ correlation at the reference redshift  \citep[here $z=2.6$, see][for this approach in considering 
the effect of the redshift evolution]{Arabsalmani15-2015MNRAS.446..990A}. This is only to visualize  the correlation after taking the redshift evolution  into account. We have here chosen the reference redshift  of $z=2.6$ since beyond this redshift the VZ correlation derived by \citet{Moller13-2013MNRAS.430.2680M} does not evolve. But it is all the same if we choose any other reference redshift for presenting the redshift-corrected VZ correlation.  }

%we have followed \citet{Arabsalmani15-2015MNRAS.446..990A} and have shifted the metallicities of the 19 hosts to the corresponding metallicities at $z=2.6$ using the evolution of the VZ correlation derived by \citet{Moller13-2013MNRAS.430.2680M}. %, where the correlation does not evolve any further with increasing redshift. 

  {
In studying the VZ correlation we are restricted to $z\gtrsim 1.7$ since absorption metallicity measurements for 
GRB hosts are limited to redshifts above  $\sim 1.7$ (observations of short-lasting GRBs optical afterglows  are usually limited 
to ground-based telescopes which do not allow the 
detection of  Lyman-$\alpha$ lines in the  spectra of GRBs at $z \lesssim 1.7$ due to atmospheric cut-off). 
However, metallicity measurements based on emission methods are available at lower redshifts. This allows 
us to investigate  the relation between the emission metallicity measurements and the kinematics characteristics 
of gas in GRB host galaxies at lower redshifts.

{
\citet{Kruhler15-2015A&A...581A.125K} performed a detailed study of  the correlation between emission metallicity measurement and 
the broadening of the bright emission lines  for GRB host galaxies. 
Splitting
their host sample into three redshift bins ($z<1$, $1<z<2$, and $z>2$) 
{they report a  correlation in the two lowest
redshift bins, and with essentially no redshift evolution.} In the higher
redshift bin they find no evidence for a correlation, instead they
find strong evidence for an evolution of the intercept (lower
panel of their Fig. 20).}

%In their full sample they found no strong correlation   and proposed this to be a result of the large redshift range in the sample host \citep[see the lower panel of Fig. 19 in][where no clear correlation between the two quantities can be seen]{Kruhler15-2015A&A...581A.125K}. They split  their host sample into three redshift bins ($z<1$, $1<z<2$, and $z>2$) and found a significant correlation   at $z<2$ with no strong evidence for a similar correlation at redshifts above 2 \citep[see the lower panel of Fig. 20 in][]{Kruhler15-2015A&A...581A.125K}. }

\begin{figure}
\captionsetup[subfigure]{labelformat=empty}
\begin{center}$
\begin{array}{c} 
\subfloat[]
{\includegraphics[trim = 0mm 10mm 0mm 0mm, clip, width=0.48\textwidth, angle=0]{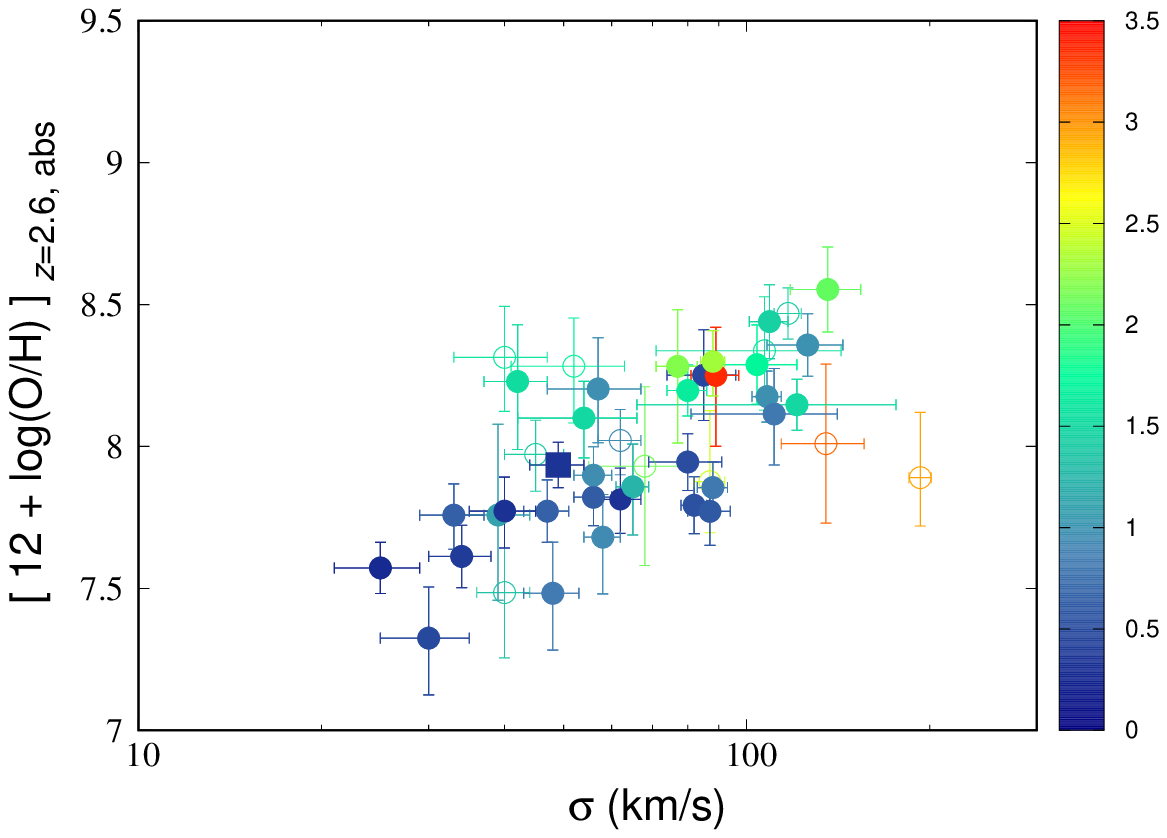}}

\cr
\subfloat[]
{\includegraphics[trim = 0mm 0mm 0mm 1.7mm, clip, width=0.48\textwidth, angle=0]{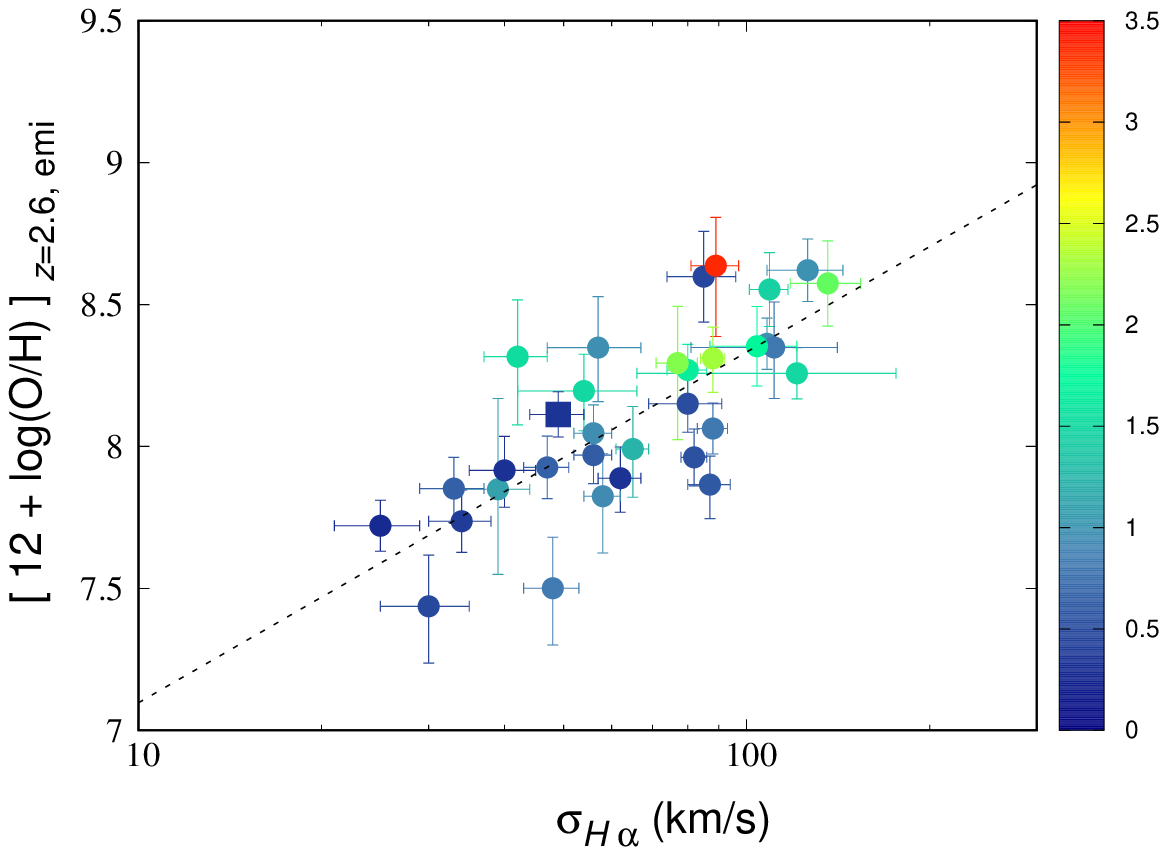}} 
\end{array}$
\end{center}
\vskip -7.7 mm
\caption[Velocity-metallicity correlation in emission]{The correlation between emission metallicity and velocity width of bright emission lines, $\sigma_{H\alpha}$ considering its redshift evolution. %Note that y-axis is different in the two panels. 
The $y$-axis %shows the measured emission metallicity, while in the other two panels it 
in both panels  shows the 
emission metallicity corrected for redshift evolution which were  set to their  corresponding values at $z=2.6$ (see Section \ref{sec:VZ} for details). In the  \textit{upper panel}  the  redshift correction of metallicities are based on the redshift 
evolution of absorption metallicities  
\citetext{\citealp[derived for QSO-DLAs by][]{Moller13-2013MNRAS.430.2680M} \citealp[and discussed for GRB-DLAs by][]{Arabsalmani15-2015MNRAS.446..990A}}.  
In the \textit {lower panel} the redshift corrections are  based on the redshift evolution of emission metallicities
obtained for general population of star forming galaxies \citep{Maiolino08-2008A&A...488..463M}.  
In the \textit{upper} 
%and \textit{middle panels}, 
panel the open circles show those  GRB hosts in our sample for which we do not have stellar mass measurements. 
We do not show these GRB hosts  in the \textit{lower panel} since redshift correction of metallicity measurements following the 
emission method requires the stellar mass measurements. 
In the \textit{lower panel} the 
dotted line  shows the best-fit correlation line.  The color-bar indicates the redshifts of the GRB hosts. 
{The host galaxy of the ultra-long GRB 130925A is marked with a square. }   
}
\label{fig:s-z}
\end{figure}

  {
We  further investigate this correlation and its redshift evolution  by shifting the metallicity measurements of all the hosts to a fixed 
redshift. We use the same sample as in \citet[][]{Kruhler15-2015A&A...581A.125K}, i.e. 43 GRB hosts  with  emission metallicity and velocity width measurements, which span  a redshift range between $z=0.28$ and $z=3.36$.
{First, %we follow the approach used in \citet{Arabsalmani15-2015MNRAS.446..990A} and 
we use the same approach  explained for Fig. \ref{fig:v-z} and shift all the metallicities to the same reference redshift of $z=2.6$ using 
the evolution of the VZ correlation in absorption.} The results are shown in the upper panel of Fig. \ref{fig:s-z}.  
The full sample clearly obeys a  tight correlation  with an intrinsic scatter of 0.16 dex. 
This is based on the assumption that metallicities in absorption and emission follow the same redshift evolution which 
may not be the case. In order to have an independent analysis from  absorption studies,  we also apply a  redshift evolution 
of the emission-line metallicities of the general  population of star-forming galaxies. 
%(in Section \ref{MZ} it will become clear that  using the metallicity evolution of the general population for GRB hosts is justified). 
{We adopt the MZ relation and its evolution up to redshift $z=3.4$ from 
\citet{Maiolino08-2008A&A...488..463M} and \citet{Troncoso14-2014A&A...563A..58T}.  For each GRB host, we calculate the offset between the measured metallicity  and the 
MZ relation at the relevant redshift. Using the calculated offsets, we  shift all the metallicities to the same reference 
redshift of   $z=2.6$. 
This can only be done for those hosts with measured stellar masses, i.e. 33 hosts out of the 43, 
as the metallicity evolution of the MZ relation  is stellar mass dependent.} 
Our results are presented in the lower panel of Fig. \ref{fig:s-z} which shows  a clear tight correlation for all the GRB hosts.  
The 10 GRB hosts without stellar mass measurements which are not shown in the lower panel  are presented with open circles in the upper  
panel of Fig. \ref{fig:s-z}. 
We  find the correlation to be (with $\rm\sigma_{H\alpha}$ in \kms unit):
\begin{eqnarray}
\label{eq:SZ}
{\rm [12+\log(O/H)]}_{z=2.6, {\rm emi}}&=& (1.24 \pm 0.19)  \log_{10}(\rm\sigma_{H\alpha}) \cr 
&+& (5.86 \pm 0.35),  
\end{eqnarray}
with an intrinsic scatter  of 0.13 dex. 
When  using either method of applying the redshift correction, the intrinsic scatter of the correlation is 
comparable with the average uncertainty in the metallicity measurements. This shows that the VZ correlation in 
emission is a significantly tight correlation.    
Note that the VZ correlation in absorption has an intrinsic  scatter of $0.4$ dex which is a about three times 
larger than the intrinsic scatter of the correlation between $\sigma_{H\alpha}$ and emission metallicity.

{
For an unbiased comparison of the absorption- versus  emission-based
redshift evolution we must compare the scatter determined from the
two methods using the same sample. Applying the absorption-based evolution
to only the  GRB hosts with known stellar masses (those shown in 
the lower panel)   we find the intrinsic scatter
to be 0.12 dex, somewhat smaller than the scatter in the lower panel. 
The emission-based redshift evolution from \citet[][]{Maiolino08-2008A&A...488..463M} 
is determined up to $z\sim 3.5$ while the absorption-based redshift evolution from 
\citet[][]{Moller13-2013MNRAS.430.2680M} is determined back to a redshift of
5.1 and is based on galaxies sampling the entire galaxy luminosity function
evenly over a wide range  \citep[see Fig. 10 in][]{Krogager17-2017arXiv170408075K}.  
The two evolution functions agree well at $z<2.6$, but at
higher redshifts they diverge. The GRB host  sample shown in the lower panel of Fig. \ref{fig:s-z} 
contains only one  host at $z>2.6$ and  hence applying the two    evolution 
functions  result in similar scatters  on the correlation.}

%To compare the  redshift evolution derived from the absorption and emission methods, we calculate the scatter in the correlation shown in the  upper panel of Fig.  \ref{fig:s-z} by including only those GRB hosts shown in the lower panel  (i.e., hosts with measured stellar mass).  \coo{We find the intrinsic scatter to be 0.12 dex, consistent but somewhat smaller than the scatter in the lower panel. Although in this analysis we are dealing with  emission metallicities, the redshift evolution derived from absorption methods \citep[from][]{Moller13-2013MNRAS.430.2680M} appears to be  consistent with the redshift evolution from emission methods \citep[from][]{Maiolino08-2008A&A...488..463M}.} 
%, but it also results in an smaller  scatter in the correlation for the GRB host sample presented in Fig. \ref{fig:s-z}. }

\section{Mass-metallicity relation}
\label{MZ}

The mass-metallicity relation  is  a fundamental scaling relation that    provides valuable 
insights into the processes which take place in formation and evolution of galaxies \citep[][]{Tremonti04-2004ApJ...613..898T, Erb06-2006ApJ...644..813E, 
Maiolino08-2008A&A...488..463M, Mannucci09-2009MNRAS.398.1915M, Zahid11-2011ApJ...730..137Z, Troncoso14-2014A&A...563A..58T}. 
GRB host galaxies with  accurate metallicity measurements obtained via  absorption profiles can in principle provide 
unique tools for studying  the MZ relation at high redshifts \citep[see][for the MZ relation at $3<z<5$ using GRB hosts]{Laskar11-2011ApJ...739....1L}. 
As a subset of star-forming galaxies, GRB hosts  are expected to follow the MZ relation of the general star-forming 
galaxy population. However, several studies have found  GRB hosts to fall below the MZ relation towards lower metallicities \citep{Stanek06-2006AcA....56..333S, Kewley07-2007AJ....133..882K, Levesque10-2010AJ....139..694L, Han10-2010A&A...514A..24H, Mannucci11-2011MNRAS.414.1263M, Graham13-2013ApJ...774..119G, 
Vergani17-2017A&A...599A.120V}. 
The typically  low metallicity of GRB host galaxies  should in principle put them on the lower mass end of the MZ relation of the general star-forming galaxy population, but still on the MZ relation. 
\citet{Mannucci11-2011MNRAS.414.1263M} suggested that the apparent low metallicity of GRB hosts compared to the general population with similar stellar masses is a consequence of the higher than average SFRs of 
GRB host galaxies. %They found GRB hosts to be fully consistent with the fundamental metallicity relation \citep[see also][]{Kocevski11-K2011ApJ...735L...8K}.
This was contradicted by \citet[][]{Graham13-2013ApJ...774..119G} who 
found  the low-metallicity preference of GRB hosts was not  driven by the anti-correlation between star formation and metallicity.

\begin{figure}
\captionsetup[subfigure]{labelformat=empty}
\begin{center}$
\begin{array}{c} \hskip -2mm

\subfloat[]

{\includegraphics[trim = 0mm 10mm 0mm 0mm, clip, width=0.50\textwidth, angle=0]{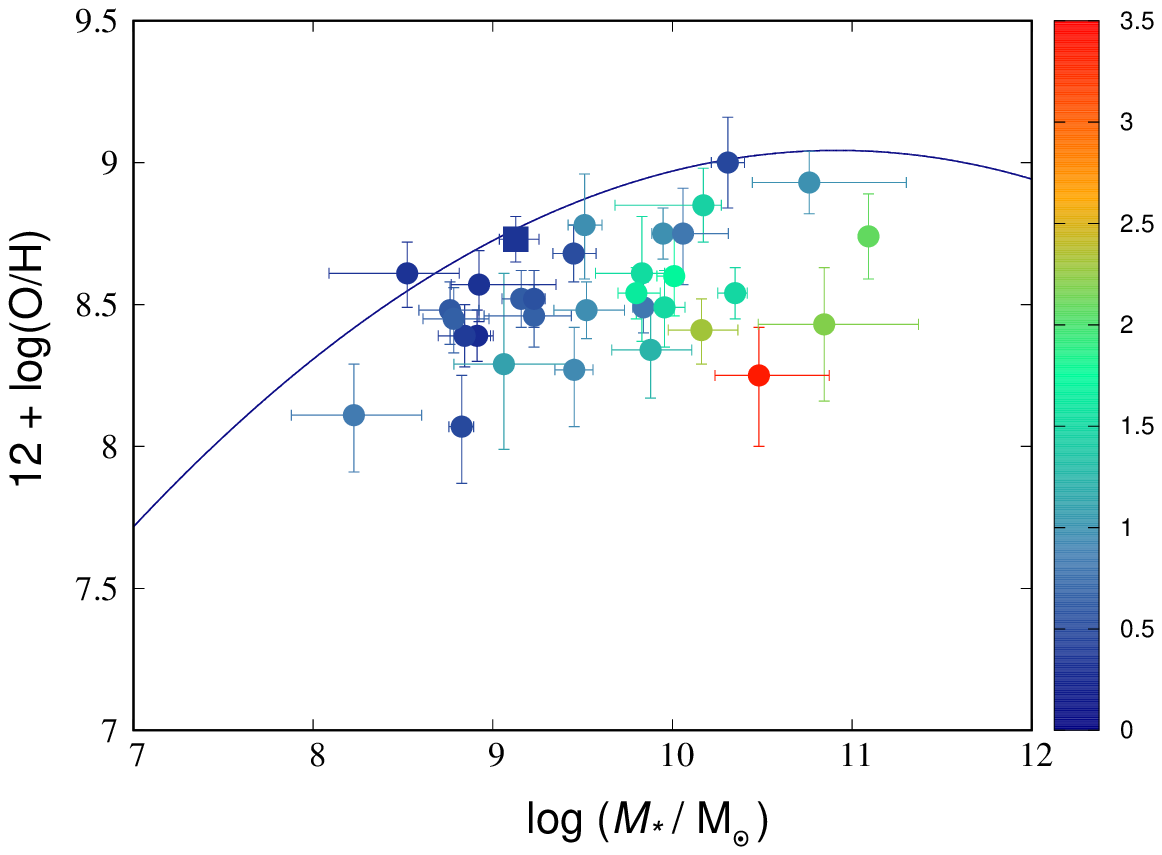}}

%& \hskip -3mm
\cr

\subfloat[]

{\includegraphics[trim = 0mm 10mm 0mm 0mm, clip, width=0.50\textwidth, angle=0]{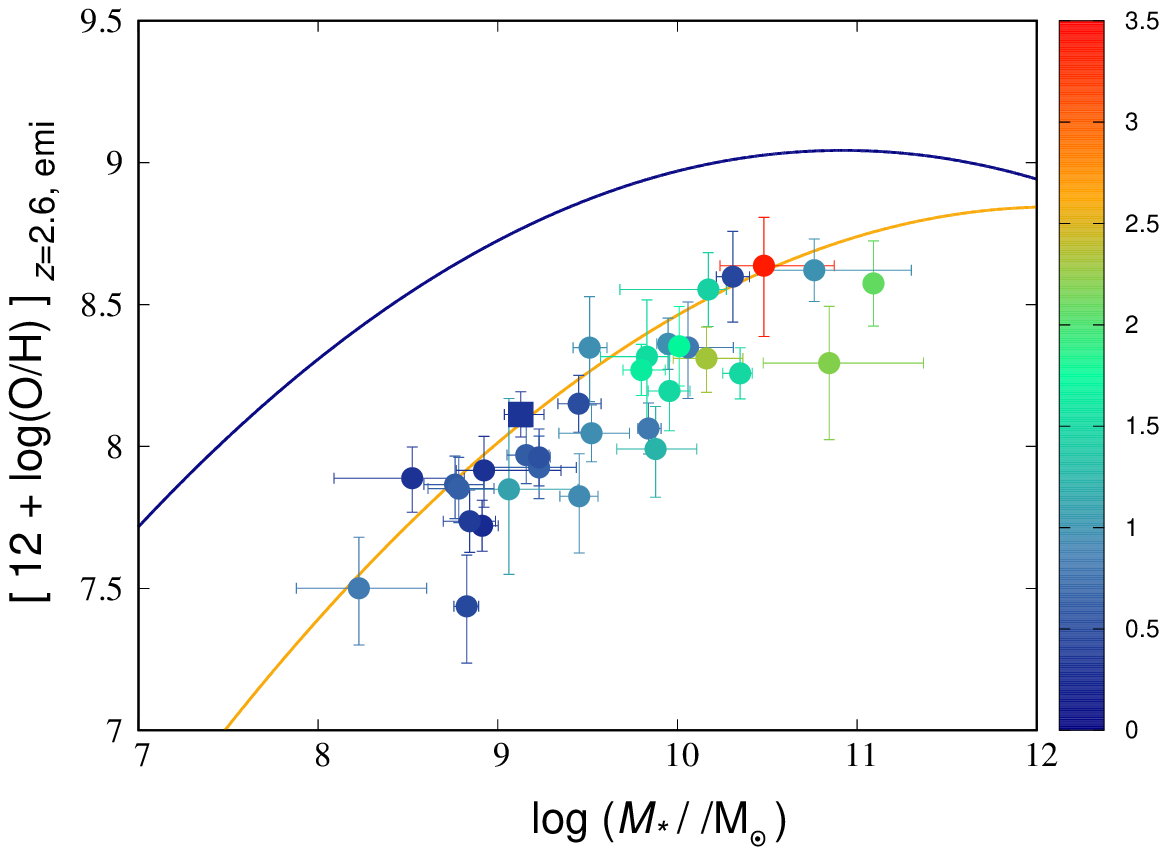}}

\cr

\subfloat[]

{\includegraphics[trim = 0mm 0mm 0mm 15mm, clip, width=0.50\textwidth, angle=0]{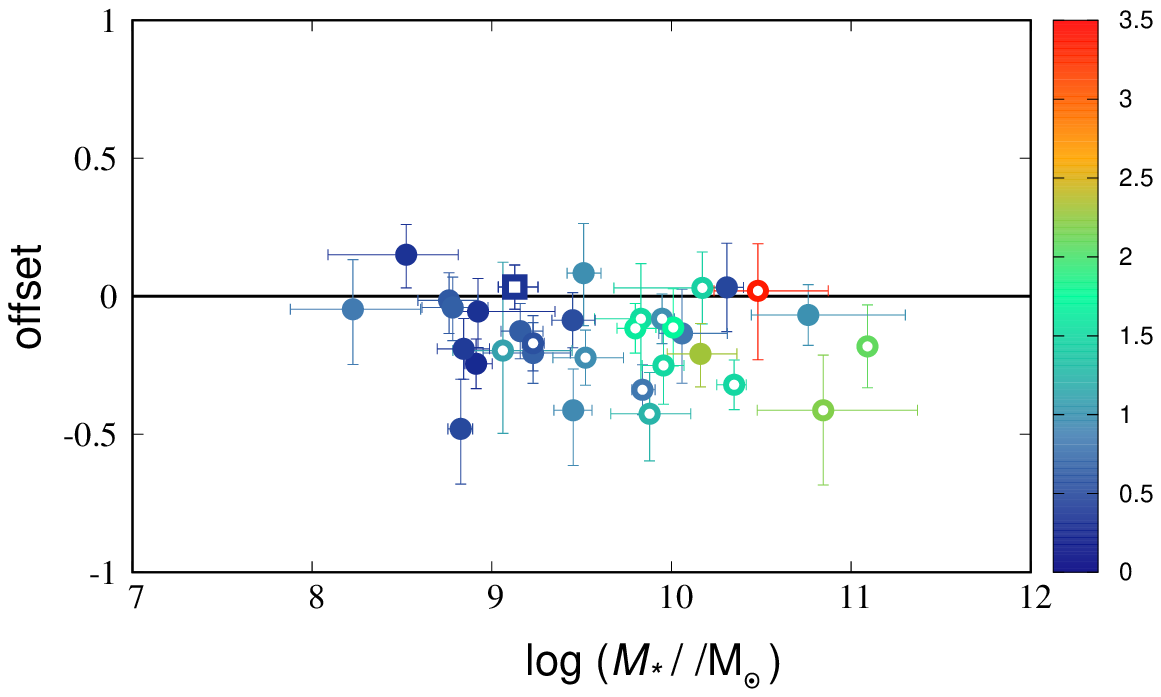}}

\end{array}$
\end{center}
\vskip -7.7 mm
\caption[Mass-metallicity relation for GRB host galaxies]{
The MZ relation for  33 GRB hosts. $y$-axis    shows  the measured emission 
metallicity in the \textit{upper panel} and its value when shifted to the reference  redshift $z=2.6$ in the \textit{middle panel}. The solid 
lines  show  the MZ relation at  $z=0$   and $z=2.6$ from  \citet{Maiolino08-2008A&A...488..463M} and \citet{Troncoso14-2014A&A...563A..58T}. 
{The host galaxy of the ultra-long GRB 130925A is marked with a square.} 
{\textit{Lower panel} shows the  offsets between the metallicity measurements of GRB hosts and the MZ relation. 
%star-forming galaxy. 
The open circles (and the open square in the case of GRB 130925A) in the lower panel  mark the  hosts of dust-obscured/dark GRBs.   
The color-bar indicates the redshifts of the GRBs and  the lines. 

}
}
\label{fig:mass-met}
\end{figure}

The existence of an MZ relation for GRB hosts and its consistency   with the MZ relation of star-forming galaxies  is still a  subject under debate.
Here we use the 33 GRB host galaxies in our sample with measured emission-line metallicities  and stellar masses 
 to study the MZ relation for GRB host galaxies and compare it to the MZ relation of the general  galaxy population. With these  33 GRB hosts we span a redshift range between $z\sim0.3$ and $z\sim3.4$ and {also cover a stellar mass range of $10^{8.2} M_{\odot}$ to $10^{11.1} M_{\odot}$}.  
%and  have significantly increased    the  sample-size compared to the samples used in previous studies of the MZ relation for GRB hosts. 
Note that the emission-line metallicity measurements for these 33 GRB hosts, taken from \citet[][]{Kruhler15-2015A&A...581A.125K}, are based on the same  diagnostics 
as used in \citet[][]{Maiolino08-2008A&A...488..463M} {and \citet{Troncoso14-2014A&A...563A..58T}}. This is important since we take  the MZ relation of the general star-forming population from these two references  for comparison with that of the GRB host sample. 
{Also, the stellar mass measurements of \citet[][]{Kruehler2017} are obtained  using the same methods and assumptions 
\citetext{\citealp[SED-fitting with galaxy templates based on][]{Bruzual03-2003MNRAS.344.1000B}, \citealp[assuming exponentially declining star formation histories with the dust attenuation curve from][]{Calzetti00-2000ApJ...533..682C}} applied for stellar mass 
measurements in \citet[][]{Maiolino08-2008A&A...488..463M} and \citet{Troncoso14-2014A&A...563A..58T}.  
The only difference is in the assumed IMF: 
Salpeter IMF used in \citet[][]{Maiolino08-2008A&A...488..463M}  vs. Chabrier IMF   used in \citet[][]{Kruhler15-2015A&A...581A.125K}. 
We use a  factor of 1.7 (0.23 dex in stellar mass) in order  to convert the results based on the Salpeter IMF to 
the corresponding results based on Chabrier IMF \citep[as done also in][]{Troncoso14-2014A&A...563A..58T}. Therefore, all the studies presented in this paper are based on assuming a Chabrier IMF.}

The upper panel in Fig. \ref{fig:mass-met} shows the MZ relation for our GRB host sample compared to the MZ relation of the general star-forming galaxy population \citep[][]{Tremonti04-2004ApJ...613..898T, Maiolino08-2008A&A...488..463M}. 
At first glance, the GRB hosts clearly appear to fall well below the MZ relation of local star-forming galaxies (at $z=0$). But as expected, GRB hosts with higher redshifts have larger deviations from the local MZ relation. 
{In order to check the  consistency of GRB hosts with the  evolving MZ relation of the general population it is appropriate to do the comparison 
in  a given redshift bin and check if all the galaxies in that bin match the MZ relation at that redshift. But since the small number of the 
GRB sample do not allow such a comparison,  we instead  plot 
our sample hosts with their metallicities shifted to a reference redshift  
 in the MZ plane and compare 
them  with the MZ relation of the general star-forming galaxy population at a reference redshift. 
As explained in previous section, for each GRB host we calculate the  offset between the  GRB host and the MZ relation of the general population at 
the GRB redshift (see the lower panel of Fig. \ref{fig:mass-met}). In order to visualize the effect of the redshift evolution and to ease the comparison we place all the GRB hosts in the MZ plane 
with the quantified offsets from the MZ relation at a reference redshift. The value of this reference redshift  has no effect on the results. }
In order to be consistent with our analysis presented in Section \ref{sec:VZ}, we choose the  reference redshift of $2.6$. 
%Since the metallicity measurements are based on emission methods, we  apply the  emission-based redshift evolution from \cite[][]{Maiolino08-2008A&A...488..463M} to shift the metallicities. 

%We should mention that this redshift evolution is determined  based  on  galaxies at $z\lesssim 3.5$ with stellar masses larger than $\sim 10^{9.5} M_{\odot}$. Thus for the GRB hosts with lower stellar masses we actually use the extrapolation of this evolution function to the lower stellar mass range.   

The middle  panel of Fig. \ref{fig:mass-met} shows our GRB host sample with
metallicities set at $z=2.6$. 
{In this plot the  GRB hosts appear to track the MZ
relation of the general population of star-forming galaxies but with an offset towards lower metallicities. 
%Note that we only quantify the offset between the GRB hosts and the MZ relation of the general star-forming galaxy population. 
%The MZ relation  of the general population is  totally independent from our GRB host sample and  using it in our analysis does not  imply that GRB hosts  should follow it. 
The lower panel of Fig. \ref{fig:mass-met} shows the  offsets between the GRB hosts and the MZ relation of the general population. 
%In order to quantify this result we determine the average offset between the GRB hosts and the MZ relation of the general population. 
We find the average offset to be  -0.15 $\pm$ 0.15 dex. 
This is in a general agreement  with the previous studies finding GRB host galaxies below the MZ relation of the general population. 
But the offset between GRB host galaxies and the MZ relation is relatively small \citep[see also][]{Japelj16-2016A&A...590A.129J}, in contradiction with  
several studies that find  GRB host  galaxies  to fall far below the MZ relation of the general star-forming galaxy population \citep[e.g.,][]{Levesque10-2010AJ....139..694L, Han10-2010A&A...514A..24H, Graham13-2013ApJ...774..119G, Vergani17-2017A&A...599A.120V}. 
We find the  offset to be comparable with the scatter on the MZ relation of the general population.
Also, the average error-bar on metallicity measurements for our GRB sample is 0.134 $\pm$ 0.002 dex which is comparable with the offset values of 0.148 dex. Therefore it is hard to decide  whether this offset is due to systematic effects or  the nature of GRB host galaxies.    
The intrinsic properties of GRB host galaxies, such as higher specific star formation rates and star formation densities \citep[e.g.,][]{Kelly14-2014ApJ...789...23K, Perley15-2015ApJ...801..102P}  could 
 lead to a similar trend in the MZ plane. It is known  that at fixed stellar masses, nearby galaxies with higher gas fractions 
typically possess lower oxygen abundances \citep[][]{Hughes13-2013A&A...550A.115H}. So possible higher gas fractions in GRB hosts (consistent with large N(H{\sc i}) values measured from the Lyman-$\alpha$ lines in GRB afterglows) could cause an offset towards lower metallicities. 
If GRB host galaxies indeed have larger outflows (see Section \ref{sec:VSM}), they would also tend to 
show lower metallicities compared to the field galaxies with similar stellar masses. 
%In addition, observational biases (which result in finding less  GRBs in dustier environments) could produce the slight offset towards lower metallicities   since dustier galaxies tend to have higher metallicities.  
%With the current data it is hard to distinguish the source of the offset such effects from the   
Systematic effects in metallicity and stellar mass measurements on the other hand could  partially be responsible for the  trend of finding GRB hosts  with an offset compared to the field galaxies on the MZ plane. 
}

The effects from observational biases (that could result in finding fewer  GRBs in dustier environments) can be addressed through 
the host galaxies of dust-obscured/dark GRBs. Such biases  may result in finding more GRBs in   galaxies with low to intermediate stellar masses 
(dustier galaxies tend to have higher stellar masses). 
This  should place the  GRB hosts at the lower mass end of the MZ relation, but is not expected    
to affect the position of the GRB-selected galaxies with respect to the MZ relation of the general population. 
In order to check the significance of such biases in our results, we consider  the dust-obscured/dark GRBs in our sample 
\citep[based on][]{Greiner11-2011AIPC.1358..121G, Perley13-2013ApJ...778..128P, Kruhler15-2015A&A...581A.125K, Perley16-2016ApJ...817....7P}
separately and check whether they show a different trend on the MZ plane compared to the whole 
sample. 
%It is the general belief that the host galaxies of dark GRBs have higher metallicities since dustier galaxies tend to have larger metallicities.  It is important to check whether dark GRB hosts behave differently from galaxies with similar metallicities and stellar masses. 
In the lower panel of Fig. \ref{fig:mass-met}  dust-obscured/dark GRB hosts are marked with the open circles. As expected,  there is no clear difference  between the dust-obscured/dark hosts and the full sample on the MZ plane. We in fact find that the  average offset of 
the dust-obscured/dark GRB hosts from the MZ relation of the general population is -0.18 $\pm$ 0.14 dex, consistent with the  
average offset of our full sample hosts towards lower metallicities  (-0.15 $\pm$ 0.15 dex).  
This  confirms that including dust-obscured/dark GRBs in our sample does not affect the scaling relations of the GRB hosts, but instead 
allows us to have better statistics and  to sample  larger ranges of galaxy properties which are critical in studying the scaling relations.  
{Also note  that  the host galaxies of the two ultra-long GRBs (GRB 111209A and GRB 130425) appear to fallow the scaling relations of the GRB host 
sample (see figures  \ref{fig:MS}, \ref{fig:VS}, \ref{fig:s-z}, and \ref{fig:mass-met}).}  
}

%Our results are in contradiction with several studies that find  GRB host galaxies  to fall far below the MZ relation of the general star-forming galaxy population.

%\citep[see also][]{Japelj16-2016A&A...590A.129J} 
{In the most recent study, \citet{Vergani17-2017A&A...599A.120V} uses a sample of 
21 GRB host galaxies at $z<2$. At low stellar masses ($M_* < 10 ^{9.5} M_{\odot}$) 
they report an agreement between GRB hosts and   the MZ relation of the general population 
\citep[with a similar sample to that of][and similar conclusions]{Japelj16-2016A&A...590A.129J}, 
but they find  GRB hosts with $M_* > 10 ^{10.0} M_{\odot}$ to be considerably offset from the  MZ relation. 
In order to explore the source of the discrepancy in between  our results and the findings of 
\citet[][]{Vergani17-2017A&A...599A.120V}, we cross check our sample with their  six GRB hosts   
with $M_* > 10 ^{10.0} M_{\odot}$. Three of these hosts are included in our analysis of the MZ 
relation. For the other three we do not have emission-line metallicity measurements from   
\citet[][]{Kruhler15-2015A&A...581A.125K} (see Section \ref{sec:sample}), but the stellar masses 
are reported for two of them by \citet[][]{Kruehler2017}. It appears that for all the five hosts  
the stellar masses used in \citet[][]{Vergani17-2017A&A...599A.120V}  are systematically larger 
(by in average 0.53 dex) than  the values that we have from \citet[][]{Kruehler2017}. This can 
indeed be the cause behind the  discrepancy in between the results. 
We also notice that at lower masses their sample is dominated by GRB hosts at 
$z<1$ where they use stellar masses measured from the  same method as in Kr{\"u}hler et al. in preparation (unlike 
their $z>1$ sample where stellar masses are measured from  the  3.6 $\mu$m flux). 
Hence,  at $M_* < 10 ^{9.5} M_{\odot}$  their stellar masses are consistent with the measurements  used in this paper 
and so their  results are consistent with our findings. }

%In addition, for their sample at $z>1$ they use stellar masses which are measured based on the  3.6 $\mu$ flux while at  $z<1$ thy use stellar mass measurements which are based on SED fits. This might have added to the systematic effects and led to having   different conclusions for the two redshift ranges.}  

The consistency of the GRB hosts with the MZ relation of the general population 
encourages the use of GRB-selected galaxies (with their  available accurate metallcity measurements) for studying the MZ relation and its evolution at high redshifts ($z\gtrsim3$).   
%at high redshifts for studying  the evolution of the MZ relation.
%The consistency of GRB hosts with the MZ relation of the general star-forming galaxy population put GRB hosts in the same category of general star-forming galaxies. This makes GRB hosts obvious candidates for studying the MZ relation and its evolution at high redshifts ($z\gtrsim3$). 
In fact \citet[][]{Laskar11-2011ApJ...739....1L} used the 
GRB hosts at $z\sim 3-5$  and found evidence for the existence of the MZ relation and its continued evolution  at $z\sim 3-5$. 
However, it should be noted that the metallicity measurements of GRB hosts at high redshifts are  obtained through absorption-line methods 
and   may differ from the emission-line metallicity measurements (see the following section). 

\begin{figure}
\begin{center}\hskip -5 mm
\psfig{file=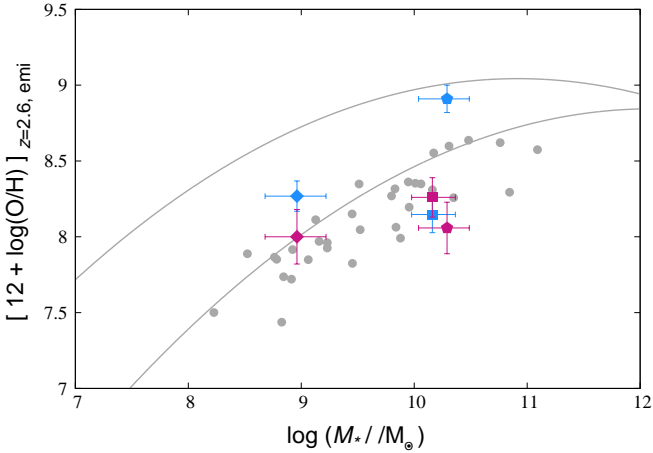,width=0.5\textwidth}
\end{center}
\caption{Three GRB hosts with measured stellar mass and absorption metallicity  on the MZ plane. 
The blue points show the 
three GRB hosts when the emission metallicities are assumed to be identical to absorption metallicity measurements and the 
magenta  points  show them with their emission  metallicities inferred form scaling relations. 
GRB 121024A is marked with squares, GRB 090323A with hexagons, and GRB 050820A with diamonds. 
The gray circles show the GRB host sample with measured emission metallicities and stellar masses. The solid  lines 
present the MZ relation of the general star-forming galaxy population at $z=0$  and $z=2.6$.  
}
\label{fig:ZXH}
\end{figure}

\section{Metallicity in absorption vs. emission}
\label{ZX}

It is necessary to confirm the consistency of the absorption metallicities 
with  metallicity measurements obtained from emission methods before using  the absorption metallicity measurements of GRB hosts to investigate the MZ relation 
at high redshifts 
\citetext{\citealp[see][for comparing emission and absorption  metallicity measurements in the sole GRB host galaxy with 
both measurements]{Friis15-2015MNRAS.451..167F}, 
\citealp[][for a similar study of a Lyman Break galaxy]{Pettini02-2002ApJ...569..742P}, 
\citealp[and][for similar studies of QSO-DLAS]{Peroux12-2012MNRAS.419.3060P, Noterdaeme12-2012A&A...540A..63N, Krogager13-2013MNRAS.433.3091K, Fynbo13-2013MNRAS.436..361F}}.

It is not clear if the metallicity measurements from absorption and emission methods should be identical as the methods of measuring the two metallicities are completely different. 
In emission, ratios of strong emission lines (like the ratio of oxygen from forbidden [O{\sc ii}] or [O{\sc iii}] lines to hydrogen   obtained from H$\alpha$ or H$\beta$ lines) are used to derive an oxygen abundance 12+log(O/H) as a  measure of the metal 
content. These methods require certain calibrations for strong-line diagnostics which are typically 
based on the physical conditions present  in low-redshift galaxies \citep[for detailed discussions see][]{Maiolino08-2008A&A...488..463M, Steidel14-2014ApJ...795..165S}.
In absorption the ratio of the column densities of metals (obtained from metal lines) to that of neutral hydrogen (obtained from Lyman-$\alpha$ line) provides a  direct and accurate 
metallicity measurement; unlike in emission {where various metallicity calibrations give rise to metallicities that differ by up to 0.8 dex for the same galaxies \citep{Kewley08-2008ApJ...681.1183K}}, absorption metallicities  do not suffer from  
calibration uncertainties and therefore are 
more reliable measurements of metal enrichment, especially at high redshifts (keeping in mind that absorption profiles provide information only in a 
narrow beam along the GRB sight-line). 
In addition, the emission and absorption profiles used in the two methods  trace different phases of gas 
 and  different regions of the galaxy. 
The absorption methods probe the metal enrichment of gas extended to the outer
most regions of the galaxy. On the other hand, the metallicity obtained from emission methods measures the metal enrichment of
the ionised gas in the star-forming  regions of the galaxy, where  star-formation  activities  have enriched the metal content of
the gas.   
Therefore it is not unexpected  if the metallicity measurements from the two methods are different. 

%Therefore, 
%even if we obtained the calibration   between the two measurement methods (which would be the relation between the two metallicity measurements in the same region),   it is likely that converting measured absorption metallicities to emission metallicities would result  in ...estimating  the metal enrichment  of the star-forming regions.
%it is likely that converting absorption metallicities to emission metallicities by assuming identical measurements from the two methods results in overestimating the emission metallicities, as presented in Fig. \ref{fig:ZXH}.
%even if both emission and absorption methods measure the same ratios  of  metal to hydrogen contents,  it is likely that  converting absorption metallicities  to emission metallicities  would result  in overestimating the metal enrichment of the star-forming regions for the galaxies with . 

In order to investigate the effect of the metallicity measurement methods in studying the MZ relation 
we use the three GRB host galaxies 
in our sample for which we have both absorption metallicity and stellar mass measurements. These are the hosts of GRB 050820A at $z=2.61$, 
 GRB 090323A at $z=3.58$, and GRB 121024A at $z=2.30$. 
The host galaxy of GRB 121024A is the sole host with measurements for both absorption and emission metallicities.  
The two metallicity measurements for this galaxy are consistent with each-other \citep[][]{Friis15-2015MNRAS.451..167F}. 
First,  we assume that the emission metallicities 
are identical with the metallicity measurements determined from absorption lines. 
The blue points in Fig. \ref{fig:ZXH} show the three hosts in the MZ plane assuming that the emission and absorption metallicities are identical. All metallicities are set to be at  $z=2.6$. 
Next, we infer  an emission metallicity for each of these hosts from their velocity width measurements and the scaling relations 
discussed in Sections \ref{sec:VSM} and \ref{sec:VZ} (the magenta points in Fig. \ref{fig:ZXH}).   {We can especially justify this approach when inferring 
an emission  metallicity based on the measurements of $\sigma_{H\alpha}$  considering the tightness 
of the VZ correlation in emission (see Section \ref{sec:VZ} for details).
GRB 121024A host galaxy  has a  $\sigma_{H\alpha}$ of $88\pm 4$ \kms. Using equation \ref{eq:SZ} (see the lower panel in Fig. \ref{fig:VS}), this leads to an inferred emission metallicity of $8.26\pm0.13$ 
(which includes  the intrinsic scatter of the used correlation, 0.13 dex, added in quadrature to the measurement error). Note that 
our  inferred emission  metallicity is  consistent with the measured value of $8.4 \pm 0.4$  reported   by 
\citet[][]{Friis15-2015MNRAS.451..167F} and $8.41^{+0.11}_{0.12}$ reported by \citet[][]{Kruhler15-2015A&A...581A.125K} for this GRB host. 
Similarly, we infer an emission metallicity of  $8.04 \pm 0.16$ for GRB 090323A host with   $\sigma_{H\alpha}=60\pm 13$ \kms. 
For the host galaxy of  GRB 050820A, we use the \vn = 300 \kms \, measurement and infer  an emission metallicity of $7.97\pm 0.18$. }
The three hosts with these inferred emission metallicities are  presented  as	 magenta  circles in Fig. \ref{fig:ZXH}. 
While acknowledging  the very small statistics,  the inferred emission  metallicities  appear to better follow 
the GRB host sample on the MZ plane  compared to when the absorption metallicities 
are assumed to be identical to the  emission metallicities. Also, in the host of GRB 090323A, the  difference between the two values is significant. 
But of course, direct measurements of emission metallicity for a handful 
number of GRB hosts with measured absorption metallicity is required  to draw robust conclusions on the relation 
between metallicity measurements from the two methods.

\section{Summary}
\label{sec:sum}

GRB host galaxies provide a unique opportunity to simultaneously study   galaxy properties obtained from absorption and emission methods. 
This includes  metallicity and  kinematics characteristics of gas, and their relations  with stellar mass,  
which provides invaluable information on galactic structure and the physical precesses leading to their formation and evolution. 
%the structure of galaxies and processes involved in their formation and evolution. 
In this paper we investigate the scaling relations between gas kinematics, metallicity, and stellar mass for a large sample 
of GRB host galaxies,  using both absorption and emission 
methods.

We  show that the velocity widths of both absorption and emission lines can be used as a  proxy of stellar mass, i.e.   
all the components contributing to the velocity widths  must be controlled 
by the gravitational potential in the galaxy. 
%We find the velocity width of low-ion absorption lines to be several times larger than that of emission lines in GRB hosts. 
We propose that the large values of \vn\, (\vn $>$ a few hundreds of \kms) can  have significant contributions  from galactic winds. 
Indeed, if galactic winds dominate the velocity width of ISM absorption lines, they appear to have 
larger velocities in GRB host galaxies compared to the general star-forming galaxy population with similar stellar masses 
\citep[see for e.g.][]{Arabsalmani17-2017arXiv170900424A}. 
This could be a result of the high SFR densities in GRB hosts. Interacting systems too could be behind such large 
velocity widths. The possible connection between mergers and GRB event have been previously pointed out in 
several cases \citep[][Roychowdhury et al. in preparation]{Chary02-2002ApJ...566..229C, Chen12-2012MNRAS.419.3039C, Arabsalmani15-2015MNRAS.454L..51A}.

We investigate the redshift evolution of the correlation between metallicity and velocity width in 
emission.  
By considering  redshift evolution, our full GRB host sample  (in a redshift range between 
0.28 and 3.36) falls on  a tight VZ correlation with an intrinsic scatter of 0.13 dex,   comparable to the uncertainty 
of the metallicity measurements. 
%We show that the redshift evolution obtained from absorption methods \citep[from][]{Moller13-2013MNRAS.430.2680M}  is  consistent with that obtained from emission methods \citep[from][]{Maiolino08-2008A&A...488..463M}.
We find the VZ correlation in emission to be significantly tighter compared to 
that in absorption.

{
We study the mass-metallicity relation of GRB host galaxies using 33 GRB hosts spanning a redshift range between $z\sim0.3$ and $z\sim3.4$ and 
a stellar mass range of $10^{8.2} M_{\odot}$ to $10^{11.1} M_{\odot}$.  By  considering the redshift evolution 
of the MZ relation, we find GRB hosts to  track  the MZ relation of the general star-forming galaxy population with 
an average  offset of  0.15$\pm$0.15 dex below the MZ relation of the general population. 
% i.e. GRB hosts follow, to within 2.5-sigma, the same MZ relation as emissionselected galaxies in the same redshift and stellar mass range. 
%In other words, to within 3-$\sigma$ GRB hosts at most have metallicities 0.06 dex lower that that of star-forming galaxies with similar stellar mass.
This offset is comparable to the scatter of the MZ relation of the general population and also to the average error-bars on metallicity measurements 
of the host sample. It is not clear if this offset is due to the systematic effects or the intrinsic properties of GRB hosts.  
%The consistency of the GRB hosts with the MZ relation of the general population   encourages the use of GRB host galaxies at high redshifts for studying  the evolution of the MZ relation. Accurate absorption metallicity measurements are available for GRB hosts at  $z \gtrsim 3$, with the  caveat that absorption metallicity measurements may differ from the emission measurements. 

We  investigate the possibility of using absorption-line metallicity measurements of GRB hosts to study the
mass-metallicity relation at high redshifts.  
Our analysis shows that  the metallicity measurements from both methods can significantly differ from each-other 
and assuming identical measurements from the two  methods may result in erroneous conclusions. 
The different metallicity estimates from  the two methods  
must be partly due to the fact that the emission and absorption profiles trace different phases of gas and
different regions of the galaxy.  

}

\section*{Acknowledgments}
M. A. would like to specially thank Thomas Kr\"uhler for very helpful discussions and also for providing stellar mass measurement 
prior to publication.   
M. A. also thanks Roberto Maiolino, Richard Ellis, Sambit Roychowdhury, Claudia Lagos, Jens Hjorth, and Bernd Husemann for helpful discussions. 
We  acknowledge the financial support from  UnivEarthS Labex program at Sorbonne Paris 
Cit\'e (ANR-10-LABX-0023 and ANR-11-IDEX-0005-02). 
The research leading to these results has received funding from the European
Research Council under the European Union's Seventh Framework Program
(FP7/2007-2013)/ERC Grant agreement no. EGGS-278202. 
AdUP acknowledge support from the Spanish research project AYA 2014-58381-P, a Ram\'on y Cajal fellowship, and a 2016 BBVA Foundation Grant for Researchers and Cultural Creators.

\begin{table}
%\begin{minipage}{100mm}
\caption{
{GRB host sample (82 GRB hosts) with GRB name and redshift  in first and second columns  (see the on-line version of the paper for 
the complete table with all values and the corresponding references listed). 
All stellar masses are taken from \citet[][]{Kruehler2017}. 
All emission-line metallicities and emission-line velocity widths are 
taken from \citet{Kruhler15-2015A&A...581A.125K}. Absorption-line velocity widths are from \citet{Arabsalmani15-2015MNRAS.446..990A} 
and this work. Absorption-line metallicities are from \citet{Savaglio06-2006NJPh....8..195S}, \citet{Ledoux09-2009A&A...506..661L}, 
\citet{Fynbo06-2006A&A...451L..47F}, \citet{Price07-2007ApJ...663L..57P}, \citet{DeCia11-2011MNRAS.412.2229D},  \citet{DElia11-2011MNRAS.418..680D},  
\citet{Thone13-2013MNRAS.428.3590T},  \citet{Savaglio12-2012MNRAS.420..627S},  \citet{DElia10-2010A&A...523A..36D}, \citet{Sparre14-2014ApJ...785..150S}, 
\citet{Delia14-2014A&A...564A..38D}, \citet{Kruhler13-2013A&A...557A..18K}, \citet{Cucchiara15-2015ApJ...804...51C}, and \citet{Friis15-2015MNRAS.451..167F}. 
}}
\vskip -5 mm
\label{Tab:sample}
\begin{center}
\begin{tabular}{lc|lc|lc}
\hline
GRB & $z$ & GRB & $z$ & GRB & $z$  \\
\hline
000926A  	&	2.0379		&     080210A  		&       2.641   &	110918A		&	0.984	 \\ 			 	
050416A		&	0.654		&     080413B		&	1.101	&	111008A  	&       5.0      \\ 		    	    	
050730A  	&       3.969   	&     080602A		&	1.820	&	111123A$^b$	&	3.1513   \\	    	    	
050820A		&	2.6147		&     080605A		&	1.641	&	111209A     	&    	0.677    \\		
050824A		&	0.828		&     080805A		&	1.505	&	111211A		&	0.479    \\ 	    	    	
050915A$^a$	&	2.528		&     081008A  		&       1.968   &	111228A$^a$	&	0.715    \\ 	    	    	
050922C 	&       2.199   	&     081109A		&	0.979	&	120118B$^a$	&	2.9428   \\ 	    	    	
051016B		&	0.936		&     081210A$^a$	&	2.063	&	120119A		&	1.729    \\ 	    	    	
051022A		&	0.806		&     081221A		&	2.259	&	120327A  	&       2.815    \\ 	    	    	
051117B		&	0.481		&     090113A		&	1.749	&	120422A		&	0.283    \\ 	    	    	
060204B		&	2.339		&     090313A  		&       3.374   &	120624B		&	2.197    \\ 	    	    	
060206A  	&       4.048   	&     090323A$^b$	&	3.583	&	120714B		&	0.399    \\ 	    	    	
060306A		&	1.560		&     090407A		&	1.448	&	120722A		&	0.959    \\ 	    	    	
060510B 	&       4.941   	&     090926A  		&       2.107   &  	120815A  	&       2.358 	 \\       	
060604A	        &	2.1355		&     090926B		&	1.243 	&	120909A  	&       3.9293	 \\       	
060719A		&	1.532		&     091018A		&	0.971   &  	121024A		&	2.301 	 \\       	
060729A		&	0.543		&     091127A		&	0.490   &  	130408A  	&       3.757 	 \\       	
060814A		&	1.922		&     100219A  		&       4.667   &  	130427A		&	0.340 	 \\       	
061021A		&	0.345		&     100418A		&	0.624   &  	130925A		&	0.348 	 \\       	
061110A$^a$	&	0.758		&     100424A$^a$	&	2.4656  &  	131103A		&	0.596 	 \\       	
061202A		&	2.254		&     100508A		&	0.520   &  	131105A		&	1.6854	 \\       	
070306A		&	1.497		&     100606A		&	1.5545  &  	131231A		&	0.643 	 \\       	
070318A		&	0.840		&     100615A		&	1.3978  &  	140213A     	&    	1.19  	 \\       	
070328A$^a$	&	2.063		&     100621A		&	0.543   &  	140301A		&	1.4155	 \\       	
070521A		&	2.087		&     100814A     	&    	1.439   &  	140430A		&	1.6019   \\		
071021A		&	2.452		&     100816A		&	0.805   &  	140506A		&	0.889 	 \\		
071031A  	&       2.692   	&     110808A		&	1.3490  &  			&		 \\	
080207A		&	2.086		&     110818A$^a$	&	3.361   & 			&	    	 \\         		  
%\hline 
%\hline
%\multicolumn{5}{c}{Emission study sample without stellar mass measurements} \\
%\hline  
%&&&&\\    
\hline	
\end{tabular}
\end{center}
$^a$ For these hosts  the emission velocity width is measured from $H\beta$ line instead of \ha line. \\
$^b$ For these hosts  the emission velocity width is measured from [O{\sc iii}] line instead of \ha line. \\
%\end{minipage}
\end{table}

\begin{table}
%\begin{minipage}{140mm}
\caption[Velocity with measurements]{  {Measurements for velocity width of  absorption lines (\vn).  
Columns 1 to 5  are GRB name, redshift, \vn, the absorption profile used for \vn\, measurement, and the smearing correction factor as defined in equation \ref{eq:r}. }
}
\vskip -3 mm
\label{Tab:v90}
\begin{center}
\begin{tabular}{llllll}
\hline
&&&&&\\
GRB & Redshift          &  $\rm\Delta v_{90}$ & low ion line & $r$ \\
    &                   & (\kms        )   &      & \\
\hline
%\hline                                              
%\multicolumn{6}{l}{Hosts with  measured \vn\, and emission line velocity width}                \\
%\hline
%&&&&&\\
091018A      &    0.971           &  146   &  SiII, 1808      &  0.15    \\
100418A     &    0.62            &  181   &  MnII, 2576      &  0.10    \\
100814A     &    1.439           &  211   &  FeII, 2600      &  0.03    \\
111209A     &    0.677           &  187   &  FeII, 2374      &  0.10   \\
111211A     &    0.4786          &  98    &  MnII, 2594      &  0.31    \\
111228A     &    0.7164          &  30    &  MnII, 2594      &  1.90    \\
%120119A     &    1.728    &   104 $\pm$  17      &  91$^b$    &  FeII, 2249      &  0.17    \\
%120119A     &    1.728    &   104 $\pm$  17      &  453$^b$   &  MnII, 2606      &  0.007    \\
120909A  &       3.9293  &       145     &   NiII, 1370 &  0.06 \\ 
121024A     &    2.301           &  437   &  MnII, 2594      &  0.01   \\
130408A  &       3.757   &       97      &   SiII, 1808 &  0.13 \\
130427A     &    0.340           &  60    &  MnII, 2576      &  0.72   \\
131231A     &    0.643           &  143   &  FeII, 2374      &  0.16    \\
140213A     &    1.19            &  151   &  FeII, 2382      &  0.14    \\
%&&&&&\\
%\hline                                              
%\multicolumn{6}{l}{3 additional hosts with  measured absorption metallicity}                \\
%\hline
%&&&&&\\
\hline
\end{tabular}
\end{center}
%\vskip 2 mm

\end{table}

\appendix 
\section{}
\label{app}
\subsection{Intrinsic scatter of the correlation}
Throughout this paper we  investigate the scaling relations between the GRB hosts 
properties in the form of linear correlations. We explain  here the method used for obtaining the correlation parameters. 
 
We basically need to find out the linear correlation between the two measurable  quantities, $y$ and $x$, 
in the form of $y=a+bx$, using a data set containing $N$ data points with measured values of $x_i$ and $y_i$ for the $i$th point. 
In some cases, the measurement errors of data points are  non-symmetric and a Monte Carlo Method  should be used  to obtain the  
best fit parameters  for the correlation. However, investigating the effect of the asymmetry of the 
error-bars, we find it to be ignorable. One reason for this is the negligible asymmetry of the error-bars 
and the other is the dominating effect of the intrinsic scatter  of the correlations, $\sigma_{{scatter}}$ (discussed below), compared to the 
measurement error-bars. Therefore, we use the standard least square method, assuming the measurement errors  of each point to be the average 
of the lower and upper measurement errors of that point. But of course we include the intrinsic scatter of the correlation as a free parameter 
in the $\chi^2$ by adding it up to the measurement error of each point in the quadratic form \citep[see][]{Moller13-2013MNRAS.430.2680M}. The $\chi^2$ then will be  
\begin{eqnarray}
\label{chi2}
\rm\chi^2 = \sum_{i=1}^{N} \frac{(a + b x_i  - y_i)^2}{\sigma_{y,i}^2+b^2 \sigma_{x,i}^2+\sigma_{\rm {scatter}}^2},
\end{eqnarray} 
where $\rm\sigma_{x,i}$ and $\rm\sigma_{y,i}$ are the average of the lower and upper measurement errors on $x_i$ and 
$y_i$ respectively, and  $a$, $b$, and $\sigma_{{scatter}}$ are the three free parameters.

\subsection{The three correlations}
The three correlations between $M_*$, $\sigma_{H\alpha}$, and $\Delta v_{90}$ are not independent 
from each other and hence they get defined by four parameters:
\begin{eqnarray}
\label{eq:wolfi}
&&M_* =a + b \, \sigma_{H\alpha},\cr
&&M_* = c+ d \, \Delta v_{90}.
\end{eqnarray} 
The two correlations in equation \ref{eq:wolfi} automatically define the correlation between $\sigma_{H\alpha}$ and $\Delta v_{90}$. 
In order to find the best fits for the four parameters we use all the three sets of data points: $n_i$ pairs of $(M_*,\sigma_{H\alpha})$, 
$n_j$ pairs of  $(M_*,\Delta v_{90})$, and $n_k$ pairs of $(\sigma_{H\alpha}-\Delta v_{90})$. Some of the data points are shared 
between the three sets. 
In order to do a  $\chi^2$ minimization that takes into account all three correlations  and the sharing of data points,we  solve a matrix optimization 
as follows. We write the three correlations for all the data points:
%For each data point we can  write one equation: 
\begin{eqnarray}
&&a + b \, \sigma_{H\alpha,i} = M_{*,i},\cr
&&c + d \, \Delta v_{90,j} = M_{*,j},\cr
&&a + b \, \sigma_{H\alpha,k} - c - d \, \Delta v_{90,k} = 0, 
\end{eqnarray}
where $i= 1,\dots,n_i$, $j=n_i+1, ..., n_i+n_j$,  and $k=n_j+1, ..., n_j+n_k$. 
To solve the equations, we write them  as a matrix equation: A .  p = M, 
where A is the matrix 
\[
\begin{bmatrix}
1 &   \rm\sigma_{H\alpha,i}  &    0  &   0 \\
0    &     0  &     1  & \rm\Delta v_{90,k}   \\   
1   &  \rm\sigma_{H\alpha,k} &   -1   &  \rm-\Delta v_{90,k}
\end{bmatrix}
\]
and p is the vector   (a,b,c,d), and   M is the vector $(M_{*,i}, M_{*,j}, 0 ... 0)$ with  the
last k elements being 0. 
The dimensions of matrix A, vector p, and vector M are  4.(i+j+k),  4, and  i+j+k respectively.

In order to make this a $\chi^2$ optimization, one has to to multiply 
both sides of each equation with the appropriate weights  before solving it. The weights  are 
$[{\sigma_{M_{*,i}}^2 + \sigma_{scatter,\, M_*-\sigma_{H\alpha}}^2}]^{-0.5}$ for equations 1 to i, 
$[{\sigma_{M_{*,k}}^2 +  \sigma_{scatter,\, M_*-\Delta v_{90}}^2}]^{-0.5}$ for equation $n_{i+1}$ to $n_i+n_j$, and 
$[{(b \, \sigma_{M_{*,k}})^2 + \sigma_{scatter, \,\sigma_{H\alpha}-\Delta v_{90}}^2}]^{-0.5}$ for equation $n_{j+1}$ to $n_j+n_k$. 
%where scatter_intrinsic_sM is the intrinsic scatter of the sigma-Mstar relation, scatter_intrinsic_vM that of the v90-M relation, and scatter_intrinsic_vs that of the v90-sigma relation.
With these weights, minimizing  $| A.p-M |^2$ is equivalent to the $\chi^2$
minimization. 
%Note that the appearance of "b" in the weighting for the lastpoints. b is initially unknown, therefore one has to iterate, i.e. solve the matrix several times. 

If a point appears in two of the data sets, the weights 
have to be reduced in order to avoid counting that measurement twice. Though we find this 
not to change our results significantly. Finally, we use numpy.linalg.lstsq routine  to solve the matrix equation. 
%. How much to reduce the weights depends on the assumptions how residuals in the two plots correlate. Will big residual in the sigma-Mstar lead to the same big residual in the v90-Mstar fit? This is unlikely, since the intrinsic scatter is quite different. Or is the only correlation in the residuals those imposed by the measurement error err_mstar. This is also highly unlikely, the reality is something in between. I therefore tried both weightings, the different in the results is not noticeable.  The weighting I tried for those object that appears in both sigma-Mstar and v90-Mstar:

%option 1:
%1/sqrt( (err_mstar_i*2)**2 + scatter_intrinsic_sM**2) for equation 1 to i
%1/sqrt( (err_mstar_k*2)**2 + scatter_intrinsic_vM**2) for equation n_i+1 to n_i+n_j

%option2:
%1/sqrt( 2* (err_mstar_i**2 + scatter_intrinsic_sM**2)) for equation 1 to i
%1/sqrt( 2* (err_mstar_k**2 + scatter_intrinsic_vM**2)) for equation n_i+1 to n_i+n_j

\subsection{An example for $\Delta v_{90}$ measurement}
Here we present an example for the \vn \, measurement. The left panel of Fig. \ref{fig:v90} shows
the MnII, 2294 absorption profile in the VLT/X-shooter spectrum of GRB 121024A at  $z=2.30$. \citet{Friis15-2015MNRAS.451..167F} present 
a detailed study of the absorption profiles in this GRB spectrum by modeling the identified profile  with a multi (five) Voigt-profile components.   
  {The right panel in the figure shows the optical depth corresponding  to the MnII, 2294 line.   
The dotted lines marked with $v_5$ and 
$v_{95}$ indicate the velocities at which  5 and 95$\%$ of the total area under the optical depth spectrum is covered respectively. The shaded area shows the 90$\%$ of the area under the apparent optical depth 
spectrum.  \vn \, is defined as $v_{95}-v_{5}$.  We measure a \vn of 434 \kms for this multi-component system.}

\begin{figure*}
\captionsetup[subfigure]{labelformat=empty}
\begin{center}$
\begin{array}{cc} \hskip -2mm

\subfloat[]

{\includegraphics[trim = 0mm 0mm 0mm 0mm, clip, width=0.47\textwidth, angle=0]{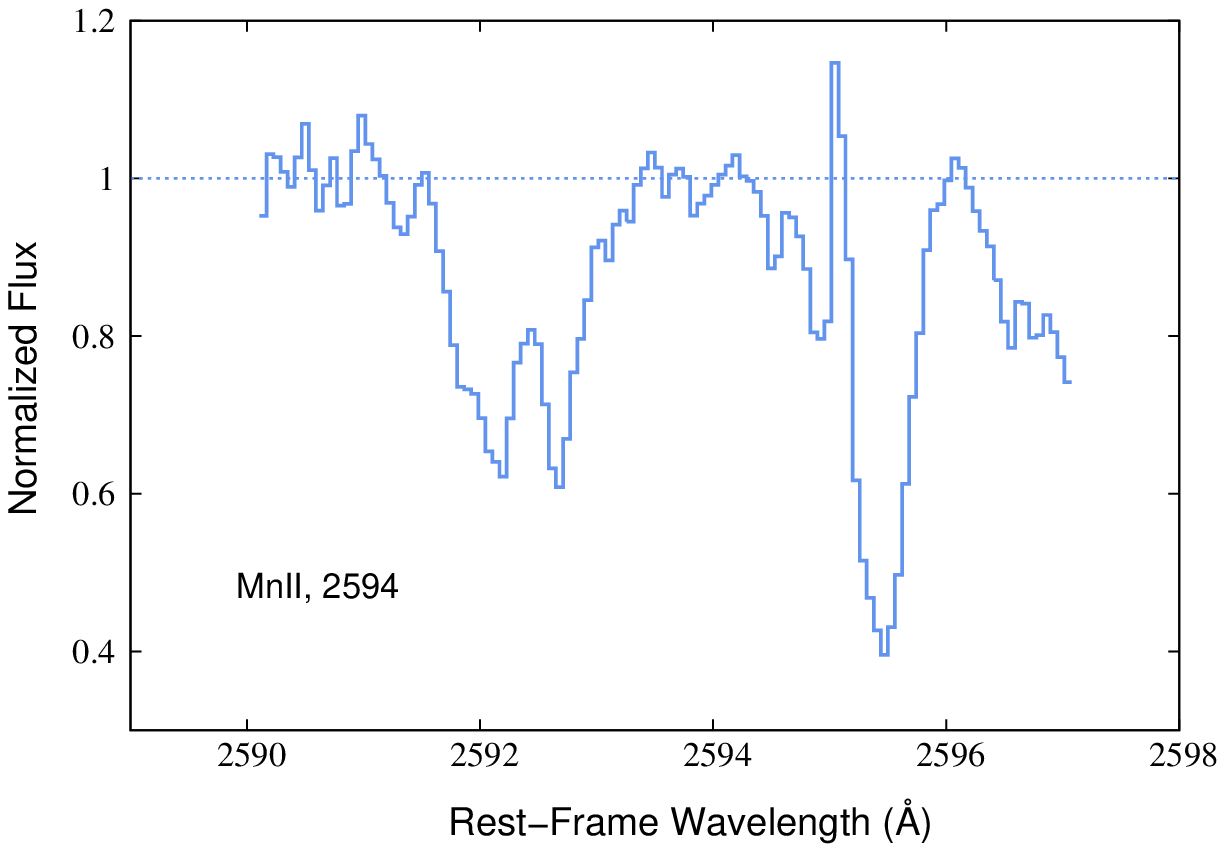}} 

& \hskip 0 mm

\subfloat[]

{\includegraphics[trim = 0mm 0mm 0mm 0mm, clip, width=0.47\textwidth, angle=0]{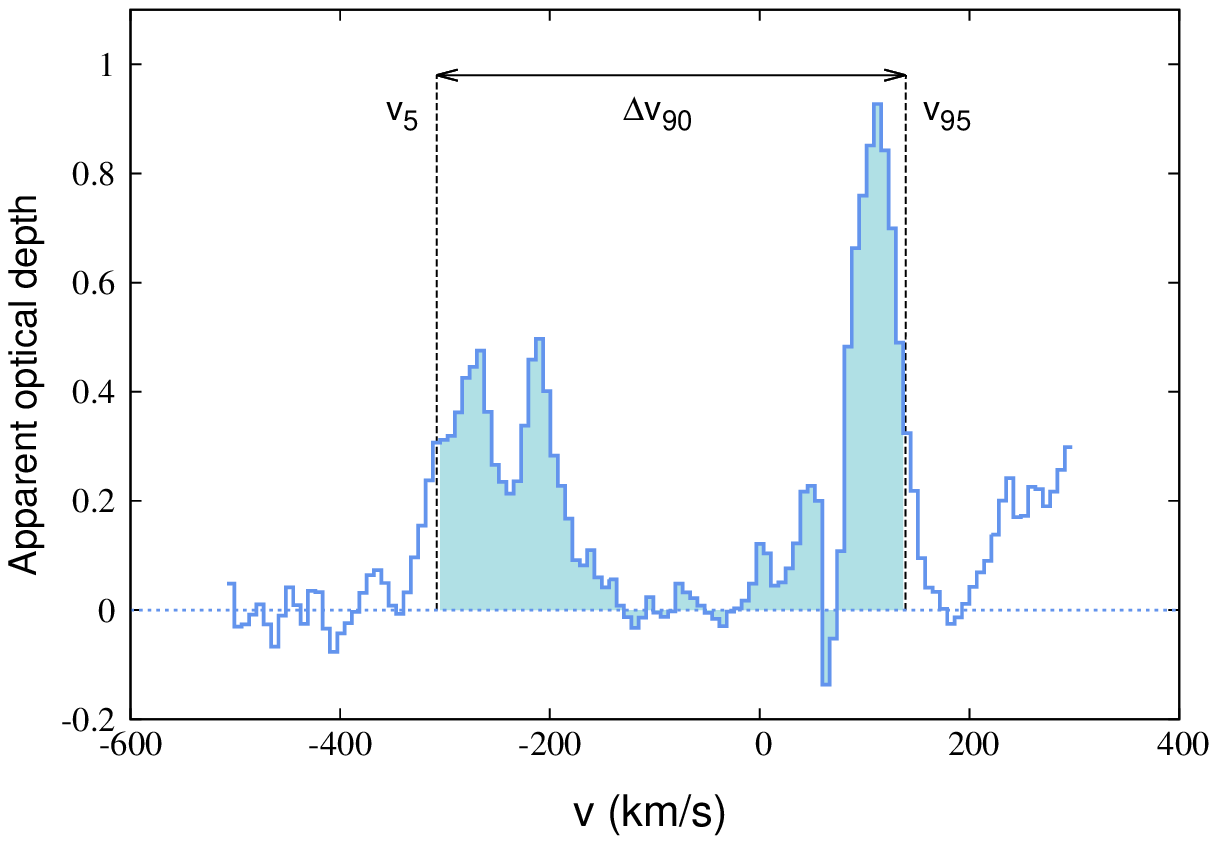}}

\end{array}$
\end{center}
\vskip -7.7 mm
\caption{
An example for \vn \, measurement from an absorption profile. { \textit {Left panel}}: The MnII, 2294 absorption profile in the VLT/X-shooter spectrum of GRB 121024A at  $z=2.30$. 
{\textit {Right panel}}: The optical depth corresponding  to the MnII, 2294 line. The dotted lines marked with $v_5$ and 
$v_{95}$ indicate the velocities at which  5 and 95$\%$ of the total area under the optical depth spectrum is covered respectively. The shaded area shows the 90$\%$ of the area under the apparent optical depth 
spectrum.  \vn \, is defined as $v_{95}-v_{5}$.   
}
\label{fig:v90}
\end{figure*}

\bibliography{adssample}
\bibliographystyle{mnras}

\end{document}